\documentclass[sigconf]{acmart}
\AtBeginDocument{%
  \providecommand\BibTeX{{%
    \normalfont B\kern-0.5em{\scshape i\kern-0.25em b}\kern-0.8em\TeX}}}

\copyrightyear{2023}
\acmYear{2023}
\setcopyright{rightsretained}
\acmConference[DIS '23]{Designing Interactive Systems Conference}{July 10--14,
2023}{Pittsburgh, PA, USA}
\acmBooktitle{Designing Interactive Systems Conference (DIS '23), July 10--14,
2023, Pittsburgh, PA, USA}
\acmDOI{10.1145/3563657.3595997}
\acmISBN{978-1-4503-9893-0/23/07}

\usepackage{enumitem}
\usepackage{hanging}
\usepackage{booktabs}
\usepackage{multirow}
\usepackage{multicol}
\usepackage{graphicx}

\newboolean{draft}
\setboolean{draft}{true} %
\ifthenelse{\boolean{draft}}{%
  \newcounter{comments}
  
  \newcommand\bl[1]{{\color{black}#1}}
  \newcommand\newt[1]{{\color{blue}#1}}
  \newcommand\strike{\bgroup\markoverwith{\textcolor{red}{\rule[0.5ex]{2pt}{0.8pt}}}\ULon}
}{

\newcommand{\bl}[1]{}
\newcommand{\newt}[1]{} %
\newcommand{\strike}{}
}

\begin{document}

\title[CoSINT]{CoSINT: Designing a Collaborative Capture the Flag Competition to Investigate Misinformation}

\author{Sukrit Venkatagiri}
\authornote{The majority of this work was completed while the author was at Virginia Tech.}
\email{sukritv@uw.edu}
\orcid{0000-0002-3888-7693}
\affiliation{%
  \institution{Center for an Informed Public, University of Washington}
  \city{Seattle}
  \state{WA}
  \country{USA}
}
\affiliation{%
  \institution{Department of Computer Science, Virginia Tech}
  \city{Arlington}
  \state{VA}
  \country{USA}
}

\author{Anirban Mukhopadhyay}
\orcid{0009-0003-0925-1084}
\affiliation{%
  \institution{Department of Computer Science, Virginia Tech}
  \city{Blacksburg}
  \state{VA}
  \country{USA}
}
\email{anirban@vt.edu}

\author{David Hicks}
\orcid{0000-0003-2660-6505}
\affiliation{%
  \institution{School of Education, Virginia Tech}
  \city{Blacksburg}
  \state{VA}
  \country{USA}
}
\email{hicks@vt.edu}

\author{Aaron Brantly}
\orcid{0000-0003-4193-3985}
\affiliation{%
  \institution{Department of Political Science, Virginia Tech}
  \city{Blacksburg}
  \state{VA}
  \country{USA}
}
\email{abrantly@vt.edu}

\author{Kurt Luther}
\orcid{0000-0003-1809-6269}
\affiliation{%
  \institution{Department of Computer Science, Virginia Tech}
  \city{Arlington}
  \state{VA}
  \country{USA}
}
\email{kluther@vt.edu}

\renewcommand{\shortauthors}{Venkatagiri et al.}

\begin{abstract}
Crowdsourced investigations shore up democratic institutions by debunking misinformation and uncovering human rights abuses. However, current crowdsourcing approaches rely on simplistic collaborative or competitive models and lack technological support, limiting their collective impact. Prior research has shown that blending elements of competition and collaboration can lead to greater performance and creativity, but crowdsourced investigations pose unique analytical and ethical challenges. In this paper, we employed a four-month-long Research through Design process to design and evaluate a novel interaction style called collaborative capture the flag competitions (CoCTFs). We instantiated this interaction style through CoSINT, a platform that enables a trained crowd to work with professional investigators to identify and investigate social media misinformation. Our mixed-methods evaluation showed that CoSINT leverages the complementary strengths of competition and collaboration, allowing a crowd to quickly identify and debunk misinformation. We also highlight tensions between competition versus collaboration and discuss implications for the design of crowdsourced investigations.
\end{abstract}

\begin{CCSXML}
<ccs2012>
   <concept>
       <concept_id>10003120.10003130.10003233</concept_id>
       <concept_desc>Human-centered computing~Collaborative and social computing systems and tools</concept_desc>
       <concept_significance>500</concept_significance>
       </concept>
   <concept>
       <concept_id>10003120.10003121.10003128</concept_id>
       <concept_desc>Human-centered computing~Interaction techniques</concept_desc>
       <concept_significance>300</concept_significance>
       </concept>
   <concept>
       <concept_id>10003120.10003121.10011748</concept_id>
       <concept_desc>Human-centered computing~Empirical studies in HCI</concept_desc>
       <concept_significance>500</concept_significance>
       </concept>
 </ccs2012>
\end{CCSXML}

\ccsdesc[500]{Human-centered computing~Collaborative and social computing systems and tools}
\ccsdesc[300]{Human-centered computing~Interaction techniques}
\ccsdesc[500]{Human-centered computing~Empirical studies in HCI}

\keywords{crowdsourcing, CTF, capture the flag, competition, collaboration, communitition, design, research through design, misinformation, OSINT}

\maketitle
\section{Introduction}

Vast amounts of publicly available data and powerful software tools have fueled the growth of crowdsourced investigations. These crowdsourced investigations have had significant real-world impact, from identifying mis- and disinformation during elections \cite{starbird2019disinformation, eip2021longfuse,prollochs2022community} to uncovering human rights abuses in war zones \cite{bellingcat2015calltoarms}, among other examples \cite{white2014digital,stowe-etal-2018-developing}. A small but growing subset of crowdsourced investigations have started to follow an ethos of analyzing only publicly available data, known as open source intelligence or OSINT \cite{dubberley2020digital}.

Crowdsourced OSINT investigations follow two main interaction approaches: collaborative and competitive \cite{belghith_compete_2022}. Examples of collaborative crowdsourced investigations include Amnesty International's Digital Verification Corps\footnote{https://citizenevidence.org/} and the University of California, Berkeley's Human Rights Center Investigations Lab.\footnote{https://humanrights.berkeley.edu/home/} The two organizations leverage a crowd of trained university students to collaboratively authenticate information on war crimes and investigate human rights abuses \cite{amnesty2020,berkeley2020}. Collaborative crowdsourced investigations have also helped to uncover misinformation \cite{prollochs2022community,Matatov2018DejaVu,juneja2022factchecking,allen2020scaling} as well as identify suspects involved in crimes \cite{venkatagiri2021crowdsolve,venkatagiri2021sedition}.

Competitive crowdsourced investigations often follow a capture the flag (CTF) model, a gamification concept borrowed from cybersecurity. In CTF competitions, teams compete against each other to capture digital ``flags'' and score points to win the game. CTFs attract thousands of participants yearly, engaging in millions of hours of collective effort \cite{noauthor_cybersecurity_2021, chapman2013picoctf}. However, most CTFs are designed to be theoretical in nature \cite{karagiannis_analysis_2020}, with little emphasis on addressing real-world problems. For example, Hacktoria's CTF seeks to introduce users to the investigative field and community of open source intelligence (OSINT) by helping them develop OSINT skills in a safe environment using simulated data \cite{hacktoria}. Unique among CTFs, TraceLabs' Search Party OSINT CTF seeks to address a real-life problem. TraceLabs' members practice their OSINT skills by searching for information about missing persons, which is then submitted to law enforcement as tips \cite{cox2018tracelabs}.

The HCI and CSCW community has extensively studied collaborative \cite{orlikowski_learning_1992,chang2017revolt,belghith_compete_2022} and competitive \cite{wen_whathack_2019,tausczik_share_2017, belghith_compete_2022,machado_collaborative_2021-1} approaches for crowdsourcing complex work, with each having unique advantages and disadvantages. Collaborations benefit from increased communication frequency and a greater sense of group cohesion; but can suffer from overly rigid hierarchies and roles, as well as a need for increased articulation work \cite{schmidt1994cooperative}. Competitions benefit by introducing a greater sense of urgency and strongly motivating novices to participate \cite{wen_whathack_2019,karagiannis_analysis_2020,carlisle2020ctf}, but can suffer from information silos and redundant effort, since each team may perform the same task and not communicate with one another \cite{tausczik_share_2017}. 

Research on data science contests \cite{tausczik_distributed_2018} and innovation contests \cite{hutter_communitition_2011} has found benefits to intentionally combining elements of collaboration and competition such as greater performance and creativity. Less work has studied collaboration and competition in crowdsourced OSINT investigations \citep[e.g.,][]{belghith_compete_2022} or sought to design hybrid approaches. Existing hybrid approaches cannot be directly applied to crowdsourced investigations because the latter pose different analytical and ethical challenges \cite{feist_synthetic_1991}. Data science contests require generating better performing models, while innovation contests require designing novel solutions. In contrast, investigations require synthesizing existing information into a coherent theory or conclusion \cite{feist_synthetic_1991, pirolli2005sensemaking}, with a greater focus on accuracy over performance and novelty. Further, improperly conducted crowdsourced investigations can result in immediate (versus delayed) harm to individuals through misidentification and harassment \citep[e.g.,][]{nhan2017digilantism, lee2020floyd, madrigal_hey_2013, venkatagiri2021crowdsolve, chang2015fleshsearch}.

In this work, we explore how to merge the complementary strengths of competition and collaboration in a new domain: crowdsourced investigations of online misinformation. We employed a four-month-long Research through Design (RtD) process \cite{zimmerman2014rtd} with 46 university students to design and evaluate a novel interaction style called \textit{collaborative capture the flag competitions} (CoCTFs). We instantiated the CoCTF concept through CoSINT, a platform that enables a trained crowd to work with professional investigators to identify and investigate social media misinformation.

To ameliorate the disadvantages of competition, such as duplication of effort and information silos, CoSINT incentivizes information sharing and collaboration between competing teams. To reduce the disadvantages of collaboration, such as increased articulation work and inflexibility, CoSINT serves as a coordinative artifact \cite{schmidt2004ordering} and gives the crowd greater agency to determine how to combine techniques and tools with scaffolding and rubrics from the field of OSINT. The competitive, gamified setting also provides additional motivation for the crowd. Finally, to augment the crowds' abilities and mitigate unwanted (unethical) behavior, they are provided with expert training and guidance.

Through our mixed-methods evaluation, we found that CoSINT enabled participants to quickly discover, archive, verify, and report on hundreds of pieces of potential misinformation on social media. Participants also said that they enjoyed using CoSINT and that it helped to better structure their workflows as they worked within their team and with other teams. Our RtD process and mixed-methods evaluation also highlighted tensions between competition versus collaboration, and in-depth versus broad investigations. 

In summary, our work makes four contributions:
\begin{enumerate}[topsep=1pt]
    \item Our paper makes a conceptual contribution by introducing collaborative capture the flag competitions (CoCTFs) to support a rapid response to investigate misinformation online. 
    \item The CoSINT platform makes a system contribution by operationalizing CoCTFs, extending the small but growing number of Research through Design studies that address misinformation \cite{arif_2018_rtd,zade2023tweettrajectory,mills2018newsthings,lovlie2022trustworthy}.
    \item Our semester-long mixed-methods evaluation with 46 students showed that CoSINT blended beneficial elements of competition and collaboration, and that participants enjoyed using CoSINT to structure their investigations.
    \item Through reflection on our design process and system evaluation \cite{zimmerman2014rtd}, we present design implications for supporting collaborative competitions in other high-stakes settings.
\end{enumerate}
\section{Related Work}
\label{sec:related_work}
To situate our work, we review prior research on approaches to investigate mis- and disinformation; systems for supporting collaborative and crowdsourced sensemaking; and ways to introduce collaboration into competitive environments.

\subsection{Open Source Intelligence Systems to Investigate Mis- and Disinformation}
Prior work identified three approaches to addressing misinformation online: agent-, message-, and interpreter-oriented \cite{wardle2017information}. Agent-oriented approaches are concerned with the specific actors that generate and spread misinformation, while receiver-oriented approaches are focused on the targets of misinformation and how receivers are affected. Each approach can also be focused on individual or a collection of: agents, messages, and receivers \cite{wardle2017information}. 

Most closely related to our work are message-oriented approaches concerned with: a) identifying content that is potential misinformation \citep[e.g.,][]{pennycook_fighting_2019} and b) verifying or refuting claims associated with or made by the content \citep[e.g.,][]{godel2021moderating}. While the growth of information and communication technologies has made misinformation more prevalent, it has also democratized access to sensitive data and powerful tools to analyze it \cite{glassman2012intelligence}. This has led to a well-established investigative field of open source intelligence (OSINT). OSINT has been widely used by media agencies \cite{bellingcat2015calltoarms}, civil society organizations \cite{noauthor_global_nodate}, governments \cite{williams_defining_2018}, and online sleuths \cite{amos_open_2022} to investigate misinformation. %

Prior work \cite{belghith_compete_2022,williams_defining_2018} identified four steps to any OSINT investigation, called the OSINT cycle: 1) discovering content; 2) verifying its provenance and determining the veracity of claims; 3) archiving the content to prevent information from being lost; and 4) reporting on the results of the investigation. Researchers have also developed software systems to support the OSINT cycle's four steps. For example, CrowdTangle~\cite{fletcher2018measuring} and Algorithm Tips~\cite{diakopoulos_towards_2021} support automated content discovery, Hoaxy~\cite{shao_hoaxy_2016} and DejaVu~\cite{Matatov2018DejaVu} support verification, the Web Archive Workbench~\cite{hswe_web_2009} supports archiving, and Birdwatch~\cite{wojcik2022birdwatch} and Maltego~\cite{hai2017maltego} support reporting. Given the growing number of OSINT systems, Abdullah et al.~\cite{abdullah_osint_2021} created OSINT Explorer, a system to help analysts determine which OSINT tool to use. 

Prior systems largely focused on individual steps in the OSINT cycle, whereas CoSINT is designed to support all four steps. CAPER \cite{aliprandi_caper_nodate} is another tool that supports all four steps to help law enforcement agents prevent organized crime. Different from CAPER, CoSINT supports investigators in investigating misinformation on social media. Although the `C' in CAPER stands for `collaborative,' it only allows sharing of \textit{files} between different law enforcement agents, and does not support collaborative \textit{work}. Unlike CAPER and most other OSINT tools, CoSINT supports virtually any number of users collaborating within and across teams on one or more investigations. Lastly, while investigations in general have been well-studied within CSCW and HCI, OSINT in particular --- despite its popularity in practice --- has garnered less attention \cite{aliprandi_caper_nodate,belghith_compete_2022, iorga2021early}. Through this work, we show that OSINT can be a valuable framework for the CSCW and HCI community.

\subsection{Systems for Supporting Collaborative and Crowdsourced Sensemaking}
Investigations involve a complex sensemaking process \cite{venkatagiri2019groundtruth,alharthi_2021,arif_closer_2017}, where investigators (e.g., journalists, historians) must collect, analyze, and make sense of disparate sources of information to arrive at a theory or conclusion presented in a final report. This sensemaking process also closely maps onto the OSINT cycle \cite{williams_defining_2018, belghith_compete_2022}, which involves similar steps (discover, archive, verify, and report). 

Investigators face challenges in adequately using OSINT tools and techniques due to limited time, personnel, access to data (e.g., reach data, follower networks, content metadata), or the data science expertise required to analyze data \cite{haughey2020misinformationbeat}. Further, many OSINT tools and techniques become obsolete because of changes in online environments and social media platforms. These challenges, coupled with the complex and dynamic nature of misinformation investigations \cite{starbird2019disinformation}, result in many investigations being under- or unexplored.

To scale up and speed up their work, investigators turn to collaboration \bl{and crowdsourcing}. Collaboration can support sensemaking by dividing discovery and verification tasks and providing diverse perspectives when analyzing data and generating reports \cite{fisher2012distributed}. Yet, collaboration also comes with coordination challenges: collaborators may be geographically distributed, have different skills and backgrounds, and have access to different data sources and tools \cite{fischer_beyond_2005,gasson2005dynamics,goyal2016effects,kang_teammate_2014}. 

Prior research on supporting collaborative sensemaking has focused on co-located teams (2--10) and crowds (30+) \cite{kerle_collaborative_2013, morris2010wesearch,vogt2011colocated,venkatagiri2021crowdsolve,hue2019sensemaking} working synchronously, or distributed crowds (30+) working asynchronously \cite{dailey2014crowdsourcerers,li2018crowdia,derthick2011collaborativewiki,fisher2012distributed,mahyar2019civicdeluge}. We extend prior work by studying how to support a distributed crowd working \textit{synchronously}. Specifically, we study a semester-long field deployment with 46 university students who conduct investigations into misinformation online in a virtually co-located classroom.

\bl{Most closely related to our work are studies of the effectiveness of crowdsourced sensemaking and fact-checking. For example, Dailey and Starbird~\cite{dailey2014crowdsourcerers} found that large, distributed crowds engaged in collective sensemaking and rumoring during crises on social media. Arif et al.~\cite{arif_closer_2017} showed that distributed crowds were able to successfully correct rumors online, and Saeed et al.~\cite{saeed2022crowdsourced} showed that crowdsourced fact-checking on Twitter frequently performed as well as professional fact-checkers. Apart from observational studies, researchers have also experimentally studied the effectiveness of crowdsourced fact-checking. Pennycook and Rand~\cite{pennycook_fighting_2019} as well as Allen et al.~\cite{allen2021scaling} found that crowdsourced trustworthiness ratings can distinguish between real and fake news sources, though Godel et al.~\cite{godel2021moderating} found that real-time crowdsourced veracity ratings performed worse than professional fact checkers. However, these prior crowdsourcing approaches studied online crowds working asynchronously or independently. Emergent collective behavior and asynchronous collaboration pose coordination challenges \cite{dailey2014crowdsourcerers, arif_closer_2017}, and aggregating independent ratings from novice crowds \citep[e.g.][]{pennycook_fighting_2019,godel2021moderating} does not allow members of the crowd to build consensus or learn from each other over time. CoSINT overcomes these limitations by leveraging trained crowds that synchronously collaborate with --- and learn from --- each other.}

Scaling up collaborative sensemaking from small teams to large crowds can amplify coordination challenges \cite{tausczik_distributed_2018}. To address these challenges, prior crowdsourcing systems have focused on dividing work into microtasks, such as collecting \cite{papoutsaki_crowdsourcing_2015}, extracting \cite{venkatagiri2019groundtruth,chan_solvent:_2018}, or schematizing \cite{crowdlines,kittur_standing_2014,chilton_cascade:_2013} data. Other projects have crowdsourced all parts of the sensemaking loop to support more complex work, such as unraveling mysteries \cite{li2018crowdia,li2019dropping} or drafting articles \cite{bernstein2010soylent,hahn2016knowledge,allen2022birds,beckett2017wikitribune}. Like this latter group of systems, CoSINT supports the entire sensemaking loop, with a focus on investigating misinformation. Still, Retelny et al. \cite{retelny2017noworkflow} found that rigid crowdsourcing workflows constrain complex and creative work. Instead, CoSINT leverages Retelny et al. and Alharthi et al.'s \cite{alharthi_2021} suggestion of including more flexible rules and roles, sharing information obtained individually with others, and fostering social interaction within the group. CoSINT allows crowd workers to choose what to collect and archive, which verification tools and techniques to use, and how to structure the final report. 

CoSINT differs from prior crowdsourcing systems and approaches in two additional ways. First, while prior work has enabled and enhanced collaborative sensemaking environments, our work begins with a competitive sensemaking environment where we introduce collaborative elements. Second, traditional crowdsourcing systems are often fixed in their functionality, relying on the developers of the system to introduce new features. CoSINT, though, handles complexity by leveraging appropriability \cite{gonzales2015appropriable, dix_designing_2007}. Roles and workflows are flexible and dynamic, and additional automated tools and crowdsourcing techniques can be integrated and configured through CoSINT's API.

\subsection{Designing Collaboration Into Competitive Environments}
Competition can benefit endeavors through providing an increased sense of urgency, immersion, and motivation \cite{yee_online_2012,park_analysis_2017}. There are several types of competitive environments, ranging from hackathons \cite{porter2017hackathons}, capture the flag competitions (CTFs) \cite{chung2014learning,carlisle2020ctf}, and innovation contests (ICs) \cite{tausczik_share_2017,lakhani2003hackers}, to games \cite{lee_effect_2014,morschheuser_designing_2017} and even self-competitions \cite{michael_race_2020}. However, while competitions such as ICs and hackathons have been used to address misinformation, one notable exception is CTFs.

In CTFs, teams compete against each other by ``capturing flags,'' either from other competitors (attack / defense-style) or from the organizers (jeopardy-style) \cite{carlisle2020ctf}. The team that captures the most flags or the highest cumulative value of flags --- where different flags are worth different numbers of points --- wins the CTF. Besides outdoor sports, CTFs are perhaps best-known in the field of cybersecurity \cite{carlisle2020ctf,mcdaniel_capture_2016,wen_whathack_2019}, requiring programming expertise to solve cryptographic puzzles, make database and network queries, or uncover exploits in operating systems. 

Prior work has shown that CTFs are best-suited for settings focused on collecting new information or uncovering new problems \cite{karagiannis_analysis_2020}. Researchers have also shown that CTFs introduce a sense of urgency and strongly motivate novices to participate \cite{mcdaniel_capture_2016,carlisle2020ctf}, while also helping novices learn through hands-on experience \cite{chung2014learning}. CoSINT leverages these features to enable a novice crowd to rapidly respond to misinformation at the message level, where a large quantity of content must be discovered and verified.

However, most CTFs are designed to be theoretical in nature \cite{karagiannis_analysis_2020}, with little emphasis on solving real-world problems. This may partly be due to the origins of some CTFs in the cybersecurity field \cite{chung2014learning,sood_chapter_2014}, where unauthorized access of real-world computer systems is illegal in the U.S. \cite{dunne1994deterring}. Still, CTFs attract tens of thousands of participants yearly, engaging in millions of hours of collective effort. For example, DEF CON's CTF attracted 3,229 teams from 2018 to 2021, each logging 276 hours of active game time \cite{noauthor_cybersecurity_2021}. Assuming three people per team, this amounts to over 2.5 million hours of effort spent on a theoretical competition that, while valuable per se for training and other other purposes, could alternatively have directly addressed a real-world problem.

Most closely related to our work, Trace Labs' \textit{Search Party OSINT CTF} is a rare instance of a real-world oriented CTF, with the goal of collecting information to help law enforcement find missing persons \cite{cox2018tracelabs}. In 2022, Trace Labs' CTF attracted 250 teams that collectively made nearly 4,000 submissions in four hours \cite{tracelabs_2022_twitter}. CoSINT builds on Trace Labs' jeopardy-style model of assigning flags of greater strategic importance more points. In Trace Labs' model, flags are independent of each other (i.e., there are no successive challenges to be completed), whereas in CoSINT, flags build on each other, starting with a discovery flag and ending with a reporting flag. Relatedly, Belghith et al.~\cite{belghith_compete_2022} studied OSINT organizations that exhibited elements of competition and collaboration (``social OSINT'') and found two other limitations of CTFs, and competitions more generally: 1) duplication of effort and 2) siloed information. These limitations are acceptable in theoretical environments, but may be less desirable in real-world investigations. Trace Labs' CTF, despite its real-world orientation, exhibits these two limitations. 

Prior work has found that introducing collaboration within a competition can help overcome these limitations. Two types of collaboration can exist in a competition: intra-team and inter-team. While intra-team collaboration naturally benefits a single team, research has shown that inter-team collaboration can be beneficial even in competitive environments \cite{tausczik_share_2017}. Analyzing 25 data science ICs on Kaggle, Tausczik and Wang found that sharing code between teams improved each individual teams' performance \cite{tausczik_share_2017}. In a design IC, Hutter et al. also found that teams who engaged in ``communitition,'' community-level collaboration among competing teams, made higher quality submissions and were more likely to win \cite{hutter_communitition_2011}. Different from prior CTFs, CoSINT includes beneficial elements of communitition found in ICs, such as sharing work products and building on others’ solutions. CoSINT also focuses on a novel domain --- crowdsourced investigations --- which poses unique analytical and ethical challenges \cite{feist_synthetic_1991}.

\section{Developing CoSINT Using Research Through Design}
Misinformation on social media is a wicked problem \cite{rittel1973dilemmas} because it is a symptom of another problem (e.g., political polarization or psychological biases), it can be interpreted and solved in many different ways (e.g., social, psychological, or technological), and solving it is identical to completely understanding it and there are no clear criteria for sufficient understanding \cite{lew_unanticipated_nodate,piccolo_2019_misinfo,montgomery_disinformation_2020}.

To address wicked problems, Zimmerman and Forlizzi \cite{zimmerman2014rtd} propose an approach to conducting research called Research through Design (RtD). RtD is the process of iteratively designing and critiquing an artifact that acts as a proposed solution to a wicked problem. Solutions to wicked problems are not right or wrong, but ``good'' or ``bad'' depending on the initial framing \cite{berger2017wicked}. Engaging in RtD enables researchers to investigate what a potential future might look like so as to reframe the wicked problem \cite{zimmerman2010analysis}. Hence, RtD has become a well-established design method in HCI \cite{berger2017wicked,zimmerman2014rtd,bardzell2016rtd,zimmerman2010analysis}.

Artifacts produced through RtD can also help inform new theory and future work, provided the process is well-documented \cite{zimmerman2010analysis, koskinen2008lab, dalsgaard2012reflective,bowers2012portfolio}, including ``detailed documentation of the actions and rationale for actions taken during the design process'' \cite{zimmerman2014rtd}. We thus describe the four phases of our RtD process for developing and evaluating the CoSINT platform: 1) frame the problem; 2) prototype workflows and interfaces; 3) clarify design goals; and 4) deploy, iterate on, and evaluate the design. %

\subsection{Phase 1: Frame the Problem}
In response to Carroll et al.'s criticism that the \textit{thing} often proceeds \textit{theory} in HCI \cite{carroll1991task}, Zimmerman and Forlizzi suggest that things in RtD should be informed by current theory and practice while spawning new theory and practice through the design and evaluation process \cite{zimmerman2014rtd}. Here, we describe our engagement with findings from prior work as well as practice. 

Recently, RtD has been applied to address misinformation by envisioning new design artifacts and interaction modalities \citep[e.g.,][] {arif_2018_rtd,zade2023tweettrajectory,mills2018newsthings,lovlie2022trustworthy,zade2023tweettrajectory}. For example, \citet{zade2023tweettrajectory} employed an RtD process to design contextual cues to inform credibility assessment on social media; while \citet{lovlie2022trustworthy} designed a tool to help readers better understand evidence and uncertainty in science journalism. The authors of this paper also have complementary experience in: 1) designing and evaluating crowdsourcing systems to support sensemaking (\bl{\cite{venkatagiri2019groundtruth, li2018crowdia,venkatagiri2021crowdsolve}}), 2) studying competitive and collaborative online communities (\bl{\cite{belghith_compete_2022,luther2013leadership,luther2009pathfinder, yu2023seditionhunters}}), and 3) participating in large-scale collaborative and competitive investigations, including hackathons and OSINT CTFs.

Through our prior experiences and engagement with the literature (Section \ref{sec:related_work}), we found three influential themes.
First, expert investigators who seek to investigate misinformation in real-time require additional resources, such as personnel \cite{haughey2020misinformationbeat}. Second, while competitive and collaborative crowdsourcing have been used to support investigators in combating misinformation \cite{belghith_compete_2022}, little work has explored how to combine both approaches in this context. Third, CTFs show promise in attracting and motivating large, novice crowds through non-monetary incentives \cite{chapman2013picoctf,carlisle2020ctf}. However, it is unclear how to adapt CTFs that are traditionally theoretical for a real-world application, as well as how to introduce elements of collaboration into a CTF. These themes led to the following research question:

\begin{quote}
\hangpara{1cm}{0} \textbf{RQ.} How can we merge the complementary benefits of competition and collaboration to provide a rapid response to investigate misinformation?
\end{quote}

\subsection{Phase 2: Prototype Workflows and Interfaces}
To inform the design of the CoSINT platform, we used existing social computing systems to piggyback prototype~\cite{grevet2015piggyback} interaction workflows and interfaces over one month with ten members of our lab acting as a small crowd. Our prototyping approach is common in classroom-oriented crowdsourcing research \citep[e.g.][]{umbelino2021prototeams,perger2014geography,dow2013pilot,zhang2017agileresearch}.

\paragraph{\textbf{Piloting Investigation Structure.}} We needed to determine how to structure the crowd's investigation into misinformation. Based on prior work \cite{lalone2018symbiotic,venkatagiri2019groundtruth,tapia2019crowdsourcing}, we chose a format that involved the crowd identifying potential misinformation, followed by verifying or refuting claims made by that content. We focused on misinformation on social media platforms given experts' need to quickly verify or refute claims made on social media before a post ``goes viral'' \cite{starbird2018journalistsrumors,venkatagiri2019groundtruth, juneja2022factchecking}. Note that \textit{potential misinformation} refers to content that has not been debunked (either verified or refuted), but appears to be falsifiable. To verify or refute claims, we used open source intelligence (OSINT) techniques, given that they are transparent and publicly accessible, and followed the OSINT cycle to structure our investigations \cite{belghith_compete_2022, williams_defining_2018,sood_chapter_2014}.

In our pilot deployments, we provided the crowd with topics to investigate, such as \textit{COVID-19, election security,} and \textit{financial misinformation}. Then, we asked the crowd to search for these topics on social media platforms, such as Twitter and Facebook, and identify potential misinformation that was recently posted and appeared to reach a wide audience (i.e., large number of shares/retweets). We prototyped different information management systems (e.g., Slack, online forms, collaborative documents, and collaborative spreadsheets). We ultimately settled on a Google Forms front-end for the crowd to quickly submit potentially misinformative social media posts in a structured manner.

\paragraph{\textbf{Piloting CTF Structure.}} Next, we needed to understand how to apply CTFs to a real-world investigative setting, given that CTFs are typically used for theoretical investigations. CTFs are often structured as individuals or teams competing against each other to capture ``flags'' and score points, with the highest-scoring team winning the competition. Capturing a flag in CTFs can represent diverse actions, e.g., identifying a piece of information, solving a cryptographic puzzle, or completing some other task. These tasks can be independent or interdependent, i.e., solving one puzzle to use the results in the next task. Many CTFs incorporate human judges to evaluate creative or subjective tasks that cannot be automatically evaluated.

\textit{Teams.} Given that our ultimate goal was to make the CTF more collaborative, we \bl{randomly assigned students} into \textsf{teams} of two or three, \bl{with each team containing at least one political science and one computer science major}. To help streamline each team's efforts, we also asked \bl{each team to nominate a team leader} and choose a unique topic to investigate.

\textit{Flags.} After consulting with the literature on OSINT and existing CTF structures, we chose four different flag types that closely map onto the OSINT cycle \cite{belghith_compete_2022, williams_defining_2018,sood_chapter_2014}: 1) \textsf{discovery flags} --- the task of identifying content that is potential misinformation, 2) \textsf{verification flags} --- verifying or refuting claims made within the content, 3) \textsf{archival flags} --- permanently archiving the discovered content and any associated information, and 4) \textsf{reporting flags} --- writing a report of the investigative process to enable transparency and reproducibility.

\textit{Points.} To make it easier to compare performance across teams and to make the evaluation more objective, we assigned \textsf{point values} to flags, with each flag type being worth 20~points. We heightened the sense of urgency and competitiveness in the CTF by creating a \textsf{leaderboard} in Google Sheets. The leaderboard displayed the cumulative points earned by each team in a graph that updated every 15~minutes.

\textit{Judges.} We introduced the concept of \textsf{judging}, where expert investigators awarded points and provided feedback on flags created by the crowd, for two reasons. First, the subjective nature of identifying potential misinformation and and verifying or refuting it necessitated expert evaluation. Second, prior work has shown that judges can increase the quality of the crowd's work by providing frequent feedback \cite{manam_wingit:_2018, dow2012shepherding}. 

Judges assessed whether a flag was relevant to the topic chosen, if the contents were falsifiable, and if the crowd worker accurately debunked claims made within the content. In our pilot deployments, the authors acted as judges due to their prior expertise with conducting such investigations. To help judges evaluate crowd workers' submissions, we asked crowd workers to include a link to a Google Doc that described their investigative process in greater detail than was included in the submission form. To provide feedback, we used Google Doc's comment feature, while teams' cumulative points were tabulated and visualized in a leaderboard using Google Sheets.

\paragraph{\textbf{Findings from Prototyping Process.}} From our prototyping process, we identified four themes that would need to be addressed in subsequent versions of our system. First, lab members said that certain types of content were more difficult to discover, and that some verifications required more time and effort than others, so they should be worth more points. Second, we found that judging flags was time consuming. We decided to make judging faster by adding more structure to the flags. 

Third, we found that some teams performed better than others due to differences in their composition (e.g., technical vs. topical expertise) and tools used (e.g., using reverse image search vs. manual searches). Fourth, we learned that most teams were often unaware of what other teams were working on. Taken together, these two findings align with prior work in other domains (data science \cite{tausczik_distributed_2018} and innovation \cite{hutter_communitition_2011}) indicating that better performance may be achieved by allowing competing teams to work with (and against) each other.

\subsection{Phase 3: Clarify Design Goals}
\label{sec:design_goals}
After settling on the general format of our CTF investigation, we engaged in a cyclic process of brainstorming and designing possible features for a collaborative CTF (CoCTF) that would overcome the limits of traditional competitions by introducing beneficial elements of collaboration. To illustrate what these features would look like, we created low-fidelity and high-fidelity interface mockups (see Appendix~\ref{appendix} for examples), and solicited feedback from lab members. 

By combining prior work and our experiences with our prototyping process, we identified two design goals: 1) support a rapid-response to investigating misinformation; and 2) give the crowd agency. However, it is challenging to design software that meets these goals in a complex setting involving a large number of people coordinating their actions \cite{gonzales2015appropriable}. 

One promising approach in CSCW for navigating this complexity is designing for \textit{appropriation} \cite{dix_designing_2007}. That is, instead of trying to understand or anticipate all of the features of a system, we can design solutions that can be used in diverse and dynamically reconfigurable ways --- thus creating more robust solutions for complex problems \cite{gonzales2015appropriable}. We instantiated our design goals by leveraging four of Dix's heuristics for software appropriation \cite{dix_designing_2007}. Dix suggests in his first heuristic that designers \textit{expose the intentions} behind the system, that is, making design assumptions and decisions explicit, and ``if they are wrong then they [can] be re-examined'' \cite{dix_designing_2007}. Along these lines, we explicitly exposed the intentions behind the CoSINT platform through our two design goals:

\paragraph{\textbf{Goal 1.}}
Our primary goal for CoSINT was to enable a rapid-response to investigating misinformation on social media. We instantiate this goal by using two of Dix's other heuristics: \textit{encourage sharing} and \textit{provide visibility}. %

\paragraph{\textbf{Goal 2.}}
Our second goal was to give the crowd agency. This is in contrast to typical crowdsourcing systems where crowds are given specific microtasks, with limited agency in how to complete them \cite{bragg2018sprout, manam_wingit:_2018}. This follows Dix's fourth heuristic to \textit{support not control} users' actions, but to provide necessary functionality so users can achieve their goals without detailed instructions. 

In the next section (Section~\ref{sec:system_description}), we describe how we designed CoSINT to meet these two goals.

\subsection{Phase 4: Deploy, Iterate on, and Evaluate the Design}
Having prototyped our designs and clarified our two design goals, we implemented them as a functional software prototype --- the CoSINT system --- and deployed it with our intended user base: a crowd of 46 students trained to investigate misinformation. 

RtD creates new situations and practices for researchers to investigate, producing gaps in behavioral theory and technical opportunity \cite{zimmerman2014rtd}. Thus, our evaluation process was continuous and sought to better understand emergent sociotechnical interactions enabled by CoSINT and opportunities for further improvement.

\paragraph{\textbf{Research and Class Setting}}
Due to the sensitive and contextual nature of investigations into misinformation, we sought crowd workers with whom we could build trust and foster accountability. In our design goals (Section \ref{sec:design_goals}), we also noted the importance of providing the crowd with more agency and allowing them to take on complex tasks through extensive training. We thus decided to deploy CoSINT in a semester-long course at our university and evaluated it using mixed methods. All authors helped to design the class and teach students OSINT investigative skills through hands-on training. This enabled a tight coupling between the skills students learned in class and how they applied then while using CoSINT. Our use of a ``class as a crowd'' is well-established, both within the domain of investigations \citep[e.g.,][]{berkeley2020,amnesty2020} and in other domains \citep[e.g.,][]{umbelino2021prototeams,perger2014geography,dow2013pilot,zhang2017agileresearch}.

For our evaluation, we required students with topical expertise in misinformation and technical expertise in testing and developing software. We thus recruited senior undergraduate and graduate students in the Computer Science and Political Science departments. The class met online due to the COVID-19 pandemic. Twice a week for approximately 90 minutes, we taught students techniques and tools in each of the four steps of the OSINT cycle. Modules for each step lasted approximately three to four weeks.%

\paragraph{\textbf{Classroom Deployment and Iteration Procedure}}
As we learned how students used the system, and analyzed their usage and feedback, we continued to iterate on the design of the software prototype for three months, until we arrived at a final system design (described in Section~\ref{sec:system_description}). Here, we briefly describe major system changes.

We demonstrated CoSINT to the class in week six and described our motivation for creating it. We also asked students to engage in an iterative, participatory design process with us. Our deployment procedure involved three steps that we repeated every two weeks (for a total of six deployments between week six and sixteen): 1) demonstrate CoSINT features and have students take part in a CTF for the entire class; 2) solicit student feedback and reflect on the CTF and system features by ourselves; 3) implement and refine new features. 

\bl{For the pilot deployment, students were placed into teams through random assignment. For the final deployment and iteration during weeks six to sixteen, students self-organized into teams of four or five members to work on the class project and compete in the CTFs. To enable group cohesion and familiarity, teams remained fixed throughout the rest of the semester. To leverage the complementary technical and domain expertise, each team was required to have least one political science and computer science major.}

\bl{Eleven research assistants and two senior researcher served as judges across the five events, with three to six individuals per event. All research team members had prior experience with open source investigations and helped with the prototyping of CoSINT. Most also had experience with participating in capture the flag competitions. For events 3 and 4, we also recruited one student from each team (different each time) to serve as a judges to gain valuable experience evaluating other teams' flags. Student judges were not allowed to evaluated their own team's flags.}

The major changes during the classroom deployment involved finding a balance of points for different flag types and for incentivizing competition versus collaboration. Based on student feedback, we modified the rubric categories and refined the user interface. Due to space restrictions, we only describe the final version of the CoSINT system in the next section.

\paragraph{\textbf{Participant Recruitment, Data Collection, and Analysis Methods}}
This study was approved by our university's IRB. The first author recruited students during an in-class guest lecture and stated that participation was voluntary. We provided consenting participants with \$20 gift cards. For the final CTF, we recognized the three top-scoring teams with prizes of \$55, \$45, and \$35. We collected both qualitative and quantitative data during and after each deployment. This included participant observation with detailed notes \cite{spradley2016participant}, notes from our weekly feedback and reflections, semi-structured interviews, and system log data. 

Twenty out of 46 students (from six out of 11 teams) participated in our study, though we report system log data on anonymized aggregate results for all teams. The median age was 21 years (range = 19--23). Fifteen (of 20) participants majored in computer science or similar fields, and four majored in political science or similar. Six identified as women, and 14 as men. None of the political science students had participated in CTFs prior to the class, while six of the computer science students had.

\begin{table*}[]
\resizebox{10cm}{!}{%
\begin{tabular}{|l|l|l|l|l|}
\hline
\textbf{Team Name}  & \textbf{Team Size} & \textbf{Participant\textsuperscript{\dag}} & \textbf{Gender} & \textbf{Degree Major} \\ \hline
OT                  & 4                  & OT-1                 & Male            & Political Science     \\ \hline
\multirow{2}{*}{SL} & \multirow{2}{*}{4} & SL-2                 & Female          & Computer Science      \\ \cline{3-5} 
                    &                    & SL-3                 & Male            & Computer Science      \\ \hline
\multirow{2}{*}{SS} & \multirow{2}{*}{4} & SS-4                 & Male            & Computer Science      \\ \cline{3-5} 
                    &                    & SS-5                 & Male            & Computer Science      \\ \hline
\multirow{4}{*}{KG} & \multirow{4}{*}{4} & KG-6                 & Female          & Political Science     \\ \cline{3-5} 
                    &                    & KG-7                 & Male            & Computer Science      \\ \cline{3-5} 
                    &                    & KG-8                 & Male            & Computer Science      \\ \cline{3-5} 
                    &                    & KG-9                 & Male            & Computer Science      \\ \hline
\multirow{4}{*}{DD} & \multirow{4}{*}{4} & DD-10                & Male            & Computer Science      \\ \cline{3-5} 
                    &                    & DD-11                & Female          & Political Science     \\ \cline{3-5} 
                    &                    & DD-12                & Male            & Computer Science      \\ \cline{3-5} 
                    &                    & DD-13                & Male            & Computer Science      \\ \hline
\multirow{3}{*}{KP} & \multirow{3}{*}{4} & KP-14                & Male            & Computer Science      \\ \cline{3-5} 
                    &                    & KP-15                & Male            & Computer Science      \\ \cline{3-5} 
                    &                    & KP-16                & Male            & Computer Science      \\ \hline
MH                  & 4                  & –                    & –               & –                     \\ \hline
TL                  & 5                  & –                    & –               & –                     \\ \hline
BF                  & 4                  & –                    & –               & –                     \\ \hline
KF                  & 5                  & –                    & –               & –                     \\ \hline
JE                  & 4                  & –                    & –               & –                     \\ \hline
\end{tabular}%
}
\Description[Table of all participants that used CoSINT]{Table of all participants that used CoSINT over the semester, their team size (varying from four to five), their self-identified gender (three out of 16 participants identified as female, the rest as male), and their degree major (three out of 16 participants were in political science, the rest in computer science). † indicates the 16 participants that we interviewed from six different teams (out of 46 students and 11 teams total).}
\caption{Table of all participants that used CoSINT over the semester, their team size, the \textsuperscript{\dag} participants that we interviewed, their self-identified gender, and their degree major.}
 \label{tab:participants}
\end{table*}

\textit{Semi-structured interviews.} The first author interviewed 16 of the 20 students who provided consent (the other four were unavailable) and took detailed notes \cite{spradley2016participant}, which were incorporated into the transcripts. The interview guide contained questions about how students worked with their team members and other teams, their perceptions of the platform and how it had changed over time, their feedback on various system components, and reflections on what they learned.

In total there were eight interviews with members from six different teams, with one to four team members in attendance for each. All participants were interviewed immediately before (KG6-9, SS-4, SS-5, OT-1) or after (SL-2, SL-3, KP14-17, DD-13) the final CTF. We interviewed DD-10 and DD-11 before and after the final CTF because they wanted to provide additional feedback. The first author recorded all interviews using Zoom and fully transcribed the recordings. Interviews ranged from 58 to 85 minutes (average = 65).

\textit{System log data.} To gain insight into how teams performed using CoSINT, we collected system log data. %
We received IRB approval to analyze anonymized, aggregated (at the team level) system log data from students who did not consent to participate in the interview portion of our study. %

\textit{Data analysis.} To analyze our data we conducted a deductive thematic analysis \cite{braun2006thematic} of the transcripts, based on themes relevant to our research questions and the various system components. These themes largely aligned with the structure of the interview guide. After downloading and fully anonymizing the system log data, we analyzed log data for all but the first event using the pandas and numpy Python libraries. We omitted the very first event from our data analysis because we made significant changes to the interface and database schema, and students would not have had sufficient time to familiarize themselves with CoSINT.

\section{The CoSINT CoCTF Platform}
\label{sec:system_description}
Here we describe the final design of the CoSINT CoCTF platform following our iterative classroom deployment process. CoSINT is designed to support experts and a team of crowd workers to rapidly respond to misinformation on social media. Teams compete against each other in an event by ``capturing'' \textsf{flags} to score \textsf{points}, and the team with the highest score at the end of the event wins the CTF. These points are shown in a \textsf{leaderboard} that dynamically updates after a team completes an action.

There are four possible actions that correspond to four flag types. Teams can 1) discover potential misinformation that makes falsifiable claims (\textsf{discovery flag}), 2) archive the content making those claims (\textsf{archival flag}), 3) verify or refute those claims and document their verification process (\textsf{verification flag}), and 4) write up a short report on their findings (\textsf{reporting flag}). Each flag type is worth a different set of points that varies according to a rubric that prioritizes higher quality flags. 

Experts specify different \textsf{narrative threads} (topics) of misinformation to investigate. Teams document a rumor or potential misinformation as an \textsf{evidence piece} that consists of one of each flag type. An evidence piece can consist of one or more of each type (except for a discovery flag, where there is a one-to-one mapping). Competing teams can also collaborate with each other by contributing flags to other teams' evidence pieces. Experts act as judges to evaluate teams' flags against the rubric and award points.

\begin{figure*}[]
\fbox{\includegraphics[width=12cm]{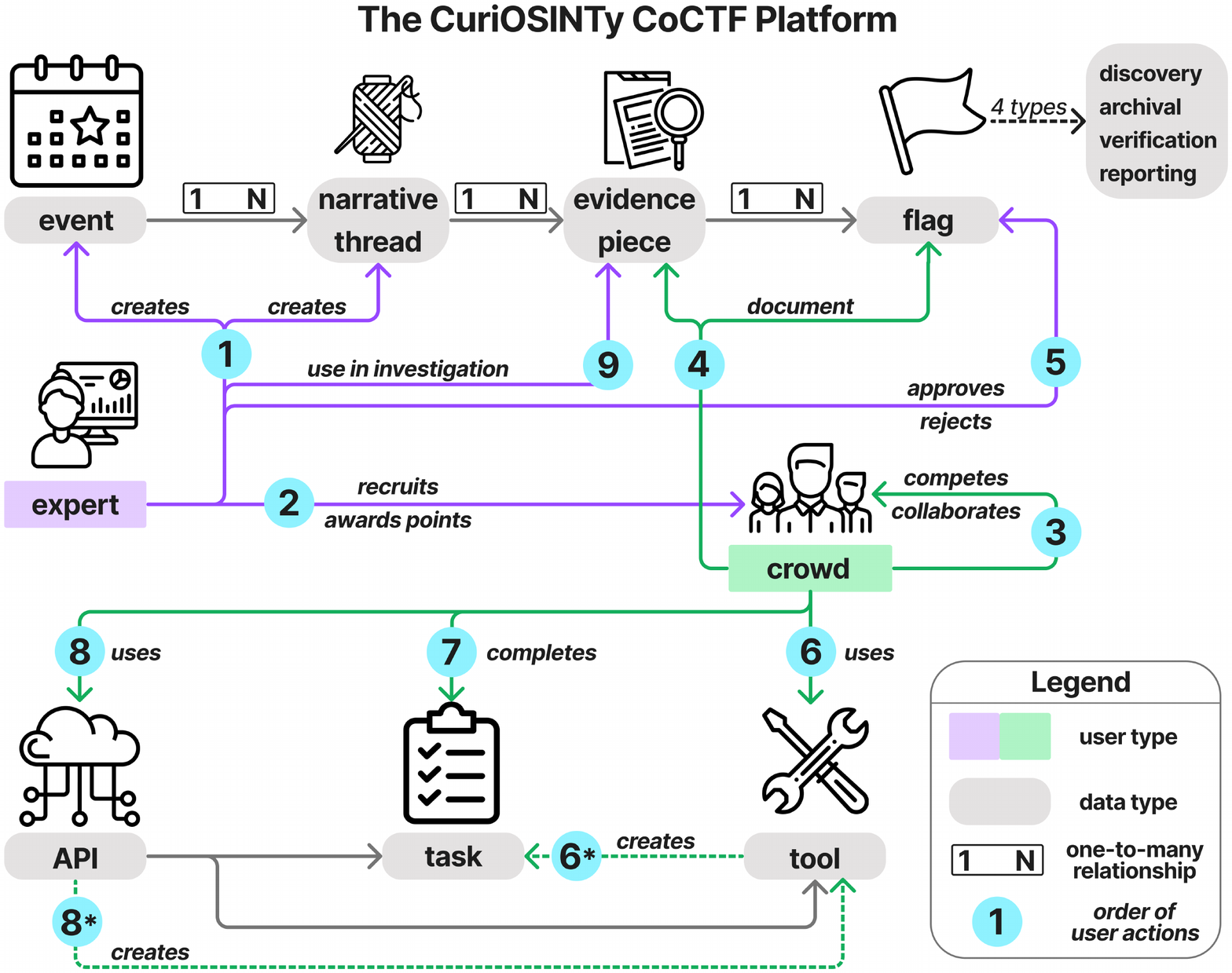}}
\caption{The structure of the CoSINT platform, described in Section \ref{sec:system_description}. Screenshots of the platform can be found in Appendix \ref{appendix}. The platform consists of events, narrative threads, evidence pieces, and flags, each hierarchically linked to the former. There are two types of users: experts and crowd workers. Experts perform actions (1), (2), (5), and (9). Crowd workers perform actions (3), (4), (6), (7), and (8). We omit further discussion of the API due to space constraints.}
\label{fig:CoSINT_system}
\Description{The CoSINT platform consists of events, narrative threads, evidence pieces, and flags, each hierarchically linked to the former. The platform also includes an API where users can create tools that allow tasks to be crowdsourced to other users. There are two general types of users: experts and crowd workers. (1) Experts create events and narrative threads. (2) Experts recruit crowd workers. (3) Crowd workers compete and collaborate with each other by (4) documenting evidence pieces and flags. (4,2) The expert approves or rejects flags and awards points to the crowd worker and their team. (6) Crowd workers can also use tools, (6*) to create tasks; and (7) complete tasks created by others. (8) Crowd workers can also use the API to (8*) create tools. Finally (9) the expert uses evidence pieces in their investigation.}
\end{figure*}

\subsection{System Description}
We now describe how the CoSINT platform (Fig. \ref{fig:CoSINT_system}) can be used to conduct an investigation into social media misinformation using a fictional scenario based on a real-life event~\footnote{https://www.amnesty.org/en/latest/news/2021/08/lebanon-one-year-on-from-beirut-explosion-authorities-shamelessly-obstruct-justice/}. Jane, the \textit{expert}, is an investigative journalist who works for a large news agency. Last night, a fire broke out in a warehouse in Beirut, Lebanon. Minutes later, two large explosions rocked the city, and were felt 150 miles away. Located several thousand miles away, Jane cannot obtain drone footage herself; and with the recentness of the event, updated satellite imagery is unavailable. However, residents of the city quickly took to social media to share what they had witnessed. Rumors about the cause of the explosion also began to spread on social media. To investigate the event and debunk false and potentially harmful claims on social media, Jane must act fast. Jane requests help from two of her co-workers (\textit{judges} in this scenario) and a group of college students (\textit{the crowd}), whom she has previously worked with. They use CoSINT to structure their work and collectively investigate the explosions in Beirut. Note: the steps below (e.g., `(1)') correspond to the numbered blue circles in Fig. \ref{fig:CoSINT_system}. 

\textbf{(1) Create Event and Narrative Threads.}
In line with our second design goal, we incorporate Dix's fourth heuristic to \textit{support not control} users' actions. CoSINT allows experts to provide some, but not complete, direction to the crowd. The expert specifies which misinformation \textsf{narratives} the crowd should track, and possibly which platforms to search. The crowd chooses how to search these platforms, which posts to investigate, and how to verify or refute a particular piece of misinformation. In this scenario, Jane first creates an \textit{event} in CoSINT, followed by several distinct \textsf{narrative threads} for teams to focus their efforts on. This includes: potential causes of the explosion, imagery of the explosion, injuries and lives lost, recent news about the port, historical information about the port, among others.

\textbf{(2) Recruit Crowd Workers and Judges.}
Through Jane's ``OSINTvestigators'' Discord group, she asks the college students to help her investigate. She also asks her colleagues, Alice and Bob, to serve as \textsf{judges}.

The students log in to the platform with their existing accounts and click on the \textsf{events} tab to access the current CoCTF event. The students form two \textsf{teams} (Gamma and Delta) and choose \textsf{team leaders}. The \textsf{leaderboard} displays all of the newly formed teams and their total points so far (currently zero).

\textbf{(3) Compete and Collaborate.}
Teams Gamma and Delta simultaneously compete against each other for points, with the highest-scoring team winning the event. However, competitions suffer from informational silos and duplication of effort. To ameliorate these limitations and in line with our first design goal, we incorporated Dix's second heuristic: \textit{encourage sharing}. Prior work studying competitions found that when communitition was encouraged (collaboration among competing teams), the individual \cite{tausczik_share_2017} and collective performance of teams was higher \cite{boudreau_open_2015, lakhani2003hackers}. CoSINT incentivizes teams to collaborate with each other in three ways. 

First, to prevent informational silos, all evidence pieces and flags are visible to all users, irrespective of team membership. Second, to enable members of Gamma and Delta to build on each others' work, any user can contribute a flag to another users' evidence piece --- scoring a collaboration point bonus, along with the base value of points for that flag. Third, there many be instances when a particular task proves too cumbersome for any one team, or if a team does not have the requisite experience but another one does. To allow users to explicitly ask for help and support others, CoSINT provides a Mechanical Turk-like requester interface. This interface, described in Steps (6,7,8), lets users create microtasks that others can complete to score additional points. 

On Discord, team leaders strategize with their team members on which \textsf{narrative threads} to address. To reduce context-switching and collaboration costs, Team Gamma (G1-4) assigns specific members to entire pieces of evidence. That is, one member will discover, verify, archive, and report on each \text{evidence piece}. Team Delta (D1-5) chooses to play off their members' strengths in discovery and verification. Delta assigns three members to discover and verify new evidence, while the other two members work on archiving the content and writing reports.

To avoid duplicate effort and further support inter-team collaboration and coordination, we incorporated Dix's third heuristic, \textit{provide visibility}, to make clear how the platform works so that the users can devise their own uses. CoSINT not only makes all information accessible to all users, but it also displays the current status of various actions \cite{schmidt1994cooperative,schmidt1994cooperative}, e.g., whether a piece of evidence is completed, or if a flag has been approved or rejected by a judge. Further, it shows users the maximum number of points they can score for each flag type and how many points a judge awarded them.

In our scenario, Team Gamma's leader, G1, views the evidence pieces that Team Delta is creating. She observes that Delta is focused on potential causes of the explosion. To avoid duplicate work, she directs her team to focus on historical information about the port. Halfway through the event, G3 learns of fertilizer storage facilities located near the explosion site, and that fertilizer is explosive. Instead of creating his own evidence piece, G3 contributes a verification flag to one of D4's evidence pieces, providing more evidence to refute a claim that the explosion was caused by a gas leak. Both G3 and D4 score points for the flags that they submitted, but also score additional points for collaborating with each other.

\textbf{(4) Document Evidence Pieces and Flags.}
To meet our second design goal of providing the crowd more agency, CoSINT incorporates scaffolding and rubrics that enable novice crowds to match expert-level performance \cite{dow2012shepherding,cook2020scaffolds}. CoSINT provides scaffolding by dividing each \textsf{narrative} into multiple \textsf{evidence pieces} focused on a particular claim. In turn, each evidence piece is divided into four different flag types (discovery, archival, verification, and reporting). However, the platform does not enforce `hard' constraints on the exact process or order of completion, but encourages high quality submissions through the point system and through a \textsf{self-assessment rubric}.

In this scenario, D5 finds a video of the explosion on Twitter. To document this, D5 clicks on the \textsf{New Evidence} button. CoSINT prompts D5 to choose an associated narrative thread (``imagery of the explosion''), specify a name, and provide the URL of the original tweet. Simultaneously, on the same page, D5 documents the \textsf{discovery flag} for discovering that evidence piece. First D5 specifies the sub-type for his discovery flag: video. Next, to promote transparency in the investigative process, D5 is required to describe how they found this content. Finally, D5 self-evaluates the quality of their discovery flag. Once D5 clicks on the \textsf{Add Evidence} button, the new \textsf{evidence piece} is created. 
 
Next, following the OSINT cycle ~\cite{dubberley2020digital}, D5 adds verification, archival, and reporting flags; and G1 adds another verification flag. A typical evidence piece consists of one of each type, but also allows multiple of each type to accommodate more complicated evidence pieces with multiple claims. Verification, archival, and reporting flags have different evaluation schemes to incentivize factors that Jane determines are important. For example, the evaluation criteria for discovery flags includes: originality, influence, and recency. More original, influential, or recent discoveries are worth more points.

\textbf{(5) Judge Flags and (2) Award Points.} 
Although Jane trusts the college students she is working with, she leverages CoSINT's judging mechanism as a first-pass filter to focus her attention on the most relevant, urgent, and accurate evidence that teams identify. Judging also gives the crowd feedback to improve the quality of flags they submit in the future. The self-assessment rubrics allow judges to evaluate flags faster, since a judge is shown the crowd worker's baseline, (hopefully) good-faith assessment of their flag.

In the scenario, Alice and Bob act as judges for the event. Judges can approve flags in any order, except for reporting flags, which can only be approved after the other three types of flags have been submitted and approved. In the first thirty minutes, teams documented several pieces of evidence and flags, which the judges begin to evaluate. 
Judges can view a user-submitted flag, the evidence piece it is part of, and the user's self-evaluation. A judge can then decide to approve or reject the flag, and modify the number of points that a team is awarded (compared to the original self-evaluation). For example, D6 submitted a verification flag where his self-evaluation totaled 600~points. However, Alice notices that some important details were missing --- such as the time of day --- awarding D6 500~points. Alice encourages D6 to submit a separate verification flag for the time of day mentioned in the original claim.

A team's points are calculated as the sum of its team members' points. If a judge rejects a flag or assigns a lower point value, a user can create a new flag with additional details and context. After D6 submits the time-of-day verification flag, he lets Alice know through the Discord group, asking her to evaluate it. Disagreements can also be clarified in a similar way.

\textbf{(6,7,8) Create and Use Tools.}
CoSINT promotes appropriability and extensibility by providing an API that supports generic task, task-response, and reward formats, similar to Human Intelligence Tasks (HITs) on Amazon Mechanical Turk. The API allows users to develop custom tools that tie into CoSINT's event structure. We omit further discussion and evaluation of the API and tools due to space constraints.

\textbf{(9) Incorporate Evidence into a Broader Investigation.}
CoSINT provides Jane with a birds-eye view of the CTF as teams and judges work. She can see all flags and evidence pieces that teams submit and filter them. She can also see the tasks that teams create using add-on tools.

This birds-eye view helps Jane direct the event and steer teams to focus on more important topics. For example, when Team Gamma discovered video footage of a nearby fertilizer storage facility, Jane realized its importance and asked them to identify the exact location where it was taken. She also asked both teams to look for other footage in nearby areas through Twitter's geotag search feature. 

After two hours, teams have collected 200~unique pieces of evidence. Jane has already looked through half of them, and created threads on Twitter debunking some social media posts that had significant traction. She is now synthesizing this evidence to write a long-form article.

\textbf{Implementation Details.}
We built CoSINT using the Python/Django web framework, a PostgreSQL database, and hosted it on Heroku.

\section{Findings}
\bl{Having described our Research through Design process and the CoSINT system, we now discuss our findings. We focus on: (1) students' evolving attitudes towards OSINT CoCTFs, followed by (2) how students reported collaborating with each other during the events, (3) their perceptions of the point system and an analysis of their actual performance based on our system log analysis, and (4) students' perceptions of judging during the events.}

\subsection{Evolving Attitudes Towards OSINT CoCTFs}

Many of the computer science students said CoSINT's format was familiar to them because they had prior experience with CTFs.
DD-12 said that the format of CoSINT --- with teams competing against each other to capture flags and score the points --- was similar to other cybersecurity CTFs he had participated in before. However, CS students pointed out four key differences they noticed. First, CoSINT had a real-world orientation (practical investigation vs. theoretical); in SL-3's words, ``instead of just reading about [investigations], we were able to perform it ourselves.'' Second, DD-12 and KG-8 said that flags were not predetermined, so there was no limit on how many points a team could score. Third, there was a different area of focus (misinformation on social media vs. cybersecurity vulnerabilities). Fourth, the time duration was shorter (60 to 90 minutes vs. several hours or days). %

In contrast with the CS students, none of the political science students was familiar with CTFs. DD-11 said that participating in gamified and fast-paced investigations was ``very much a culture shock for us'' because ``quite literally, everything is different.'' Still, DD-11 felt that the format of the investigation was advantageous for political science students because of its novel, hands-on aspect: ``[I]t isn’t just writing a paper on this topic that we’ve researched for a few weeks.'' Instead, she said the CTF taught efficiency and teamwork with a focus on addressing a real-world problem. Despite the novelty, OT-1 said that the instructions we provided were clear and that participating in the CTF was ``pretty easy once you get the hang of it.''

\subsubsection{Defining success}

All students said that one form of success was scoring the most points and winning the CTF. One student, SL-2, said she was ``very competitive in pretty much anything'' and would feel successful if she could ``find the way to most easily and effectively win within [the] bounds'' of a competition’s rules.

Students' definition of success evolved over the semester, to the point where many said that success was more than just winning the CTF. Other markers of success included finding actionable misinformation, achieving a ``flow state'' \cite{csikszentmihalyi1990flow} with their teammates, having an enjoyable experience, and learning new skills to grow as an investigator.

Not all students were equally motivated by competing to win the CTF. Although DD-10, DD-12, and OT-1 acknowledged that success meant being on top of the leaderboard, they found the most enjoyment when they worked with their team members to find ``a story or some sort of a coordinated campaign [which] much more successful than just collecting a bunch of unrelated flags'' (DD-12).

\subsection{Collaboration Styles During the CoCTFs}
We asked students how they worked within their teams and found two types of workflows that they employed: (1) assembly line and (2) free-for-all.

\subsubsection{Assembly line workflow}
In the assembly line workflow, each member focused on certain phases of the investigation, with two or more team members working on the same piece of evidence. Two teams (DD, KGB) developed an informal leadership structure with their assembly line. Team KP --- which had largely used a free-for-all workflow (see below) --- also set up an assembly line workflow for the final CTF.

DD-12 said he disliked how each team member worked independently, instead of deeply working together to complete every flag. Unlike DD-12, KP-14 enjoyed the efficiency of the assembly line:
\begin{quote}
We just were very fluid, moving very quickly, [\ldots] and we kind of spread all that work out. And, you know, I would find evidence, [KP-16] would archive it \ldots I felt really good when the flags that I submitted would actually get approved.
\end{quote}

Informal team leaders emerged over time for Teams DD and KGB. DD-12 pointed out how he and DD-10 became the \textit{de facto} team leaders and that it ``wasn’t done intentionally that way.'' DD-10 and 12's team member, DD-13, wished for a way to explicitly assign work to team members to more easily manager the assembly line workflow.

\subsubsection{Free-for-all workflow}
Different from the assembly line workflow, four teams (SS, SL, OT, KP) employed workflows that were largely ``free-for-alls'' where decisions were made in an ad hoc manner. These team members largely worked without coordination on a given evidence piece, submitting flags for each of the four phases. For example, SL-3 said, ``it was mostly a free-for-all \ldots We’d be talking about our flags and what we found. But we wouldn’t really collaborate on or delegate specific tasks.''

Another student, SS-5, believed that the free-for-all workflow was better than an assembly line workflow. He said that his team initially followed an assembly line workflow, but quickly decided to switch to working largely independently. This was because it was difficult to communicate intention and context in an assembly line workflow: ``The first person’s discovery flag doesn't communicate well to the other person trying to do the verification. `Hey, why do you actually think this is misinformation?' or `Why do you think this needs to be verified in the first place?' So you can't really just split it up purely into those stages''' (SS-5).

Apart from Team SL, three of the four teams did not have a team leader. OT-1 said his team would take on tasks at the start of each CTF without a leader assigning them. %
For Team SL, SL-3 said that SL-2 became the de facto team leader because of her strong performance over the first few CTFs.

\subsection{Point System Was Effective But Revealed New Tensions}
We found that, in line with prior work \cite{tausczik_share_2017,belghith_compete_2022}, CoSINT's point system initially promoted a competitive environment. About two thirds of students said that they enjoyed the competitive environment, while one third said that they did not. As the semester progressed, we modified the point values in line with our goal to make CoSINT more collaborative, while also taking into account students’ feedback on the relative balance of points assigned to different flag categories. However, given the fast-paced, largely competitive, and gamified structure of CoSINT, we found two key tensions over the course of the semester. The first is a tension between competition and collaboration, and the second is a tension between quantity and quality. %

\subsubsection{Performance}
In Fig.~\ref{tab:evidence_flag}, we see that for Event 1, teams submitted 227 flags across 148 evidence pieces (mean = 1.54 flags per evidence). By Event 5, teams submitted 597 flags across 228 evidence pieces (2.62 flags per evidence) --- a 70.1\% increase in flags per evidence. Despite a 163\% increase in the number of flags, the flag approval rate was similar: 78.5\% at Event 1 versus 80.4\% for Event 5. For Events 2--4, we see that approval ratings were slightly higher at approximately 90\%. This suggests that teams became more efficient at submitting more flags without a corresponding decrease in the approval rating.

\begin{table*}[]
\begin{tabular}{|l|l|l|l|l|l|}
\hline
\textbf{Type}                                                                                             & \textbf{Event 1} & \textbf{Event 2} & \textbf{Event 3} & \textbf{Event 4} & \textbf{Event 5} \\ \hline
\textbf{Total no. of flags}                                                                               & 227              & 158              & 257              & 238              & 597              \\ \hline
\textbf{No. of approved flags}                                                                            & 179              & 144              & 229              & 209              & 480              \\ \hline
\textbf{No. of rejected flags}                                                                            & 48               & 14               & 28               & 29               & 117              \\ \hline
\textbf{Flag approval rate}                                                                               & 78.9\%           & 91.1\%           & 89.1\%           & 87.8\%           & 80.4\%           \\ \hline
\textbf{No. of verification flags}                                                                        & 29               & 22               & 53               & 40               & 93               \\ \hline
\textbf{\begin{tabular}[c]{@{}l@{}}No. of verification flags \\ identifying misinformation\end{tabular}}  & N/A                & 7                & 37               & 27               & 83               \\ \hline
\textbf{\begin{tabular}[c]{@{}l@{}}Pct. of verification flags \\ identifying misinformation\end{tabular}} & N/A              & 31.82\%          & 69.81\%          & 67.50\%          & 89.25\%          \\ \hline
\textbf{Total no. of evidence}                                                                            & 148              & 97               & 112              & 114              & 228              \\ \hline
\textbf{Total no. of flags per evidence}                                                                  & 1.54             & 1.63             & 2.3              & 2.09             & 2.62             \\ \hline
\end{tabular}
\Description[The number of evidence pieces and flags created per event]{The number of evidence pieces and flags created per event. From Event 1 to 5, the number of flags created increased from 227 to 597, and the number of flags per evidence increased from 1.54 to 2.62, while flag approval rate stayed approximately the same at 80 to 90\%. From Event 2, we introduced a way to track whether a verification flag refuted the original claim (i.e., identified misinformation), and we see an increase from 31.82\% to 89.25\%.}
\caption{The number of evidence pieces and flags created per event. From Event 2, we introduced a way to track whether a verification flag refuted the original claim (i.e., identified misinformation).}
\label{tab:evidence_flag}
\end{table*}

\subsubsection{Receptiveness to the Point System}
We found that students’ receptiveness to the points system was affected by a combination of motivational factors --- both intrinsic (sense of achievement, competence, and learning) and extrinsic (monetary incentives, grade incentives).

Most students, such as KG-9 and KP-15, were extrinsically motivated by the points system to participate and develop their skills. For them, the point-based leaderboard provided direct feedback indicating whether their strategies were effective. For example, KG-9 said, ``I like being able to look at the scoreboard and be like, `Hey, we did pretty good today.’ Or sometimes we have bad days, too. And then you learn from the bad days, like, ‘Oh, maybe I should have found more misinformation.’''

Some students did not enjoy competition-based games in general. One graduate student, DD-13, said that it might be an ``an age thing'' where they were no longer motivated by competing for its own sake. Others were driven by more intrinsic motivations, such as the thrill of the hunt in conducting investigations. %
For example, KP-15 said, 
\begin{quote}
I don’t think points necessarily correlate to how good of an investigator you are. I think it also has a lot to do with your strategy and which things you focus on. I really enjoyed things that felt engaging to me, like maybe an original verification, archival, or a really good discovery.
\end{quote}

\subsubsection{Changing Incentives Can Affect Desired Outcomes}
Through the point system, we sought to value work that was of greater strategic importance (more recent, more reach, etc.), of higher quality, or required more effort to do. We found that we could encourage students to focus more (or less) on certain aspects of the investigation by changing the point values for different flags and flag categories. For example, KP-15 focused more on verifications once they noticed that verification flags were worth more points than discovery flags:

\begin{quote}During the first few CTFs, I would mainly focus on discovery. And I don’t think I’d get a whole lot of points from discovery \ldots When I did one verification, it got me as many points as it took for five discoveries. Once I noted that, I was like, `What am I doing? I should focus on verification,' because for the same amount of time I can get way more points. \end{quote}

Acting on this realization, KP-15 and his team placed second in Event 5, where 61.8\% of their points came from verification flags. In previous events, the percent of Team KP's points that came from verification ranged from 0\% in Event 1 to 54.9\% in Event 4. For Event 5, Teams KP and OT (who placed fourth) were the only two teams who received 60\% or more of their points from verification flags. On the other hand, for all other teams in Event 5, the average percent of points received from verification flags was 33\% (min. = 0\% and max.= 47.3\%). 

Interestingly, the team that placed first in Event 5, Team SL, submitted only one verification flag (which was rejected). Instead, it appears that Team SL opted to obtain the majority of their points from discovery and archival flags (40.8\% and 39.5\%, respectively). Team MH, who placed third, chose a more evenly distributed strategy, obtaining 31.5\% of points from verification, 20\% from discovery, 16\% from archival, and 8.8\% from reporting flags.

\begin{table*}[]
\addtolength{\tabcolsep}{-1.5pt}
\small
\begin{tabular}{|l|ll|ll|lll|lll|llll|l|}
\hline
\multirow{2}{*}{\textbf{\begin{tabular}[c]{@{}l@{}}Event \\ / \\ Team\end{tabular}}} & \multicolumn{2}{l|}{\textbf{Event 1}}  & \multicolumn{2}{l|}{\textbf{Event 2}}           & \multicolumn{3}{l|}{\textbf{Event 3}}                                                                                          & \multicolumn{3}{l|}{\textbf{Event 4}}                                                                                          & \multicolumn{4}{l|}{\textbf{Event 5}}                                                                                                                                                                        & \multirow{2}{*}{\textbf{\begin{tabular}[c]{@{}l@{}}Avg. \\ Rk\end{tabular}}} \\ \cline{2-15}
                                                                                     & \multicolumn{1}{l|}{Rk}         & Pts  & \multicolumn{1}{l|}{Rk}         & Pts           & \multicolumn{1}{l|}{Rk}         & \multicolumn{1}{l|}{\begin{tabular}[c]{@{}l@{}}Pct. \\ Collab.\end{tabular}} & Pts           & \multicolumn{1}{l|}{Rk}         & \multicolumn{1}{l|}{\begin{tabular}[c]{@{}l@{}}Pct. \\ Collab.\end{tabular}} & Pts           & \multicolumn{1}{l|}{Rk}         & \multicolumn{1}{l|}{\begin{tabular}[c]{@{}l@{}}Pct. \\ Tasks\end{tabular}} & \multicolumn{1}{l|}{\begin{tabular}[c]{@{}l@{}}Pct. \\ Collab.\end{tabular}} & Pts            &                                                                              \\ \hline
\textbf{OT}                                                                          & \multicolumn{1}{l|}{5}          & 1724 & \multicolumn{1}{l|}{4}          & 2753          & \multicolumn{1}{l|}{4}          & \multicolumn{1}{l|}{0\%}                                                  & 7637          & \multicolumn{1}{l|}{\textbf{1}} & \multicolumn{1}{l|}{0\%}                                                  & \textbf{8716} & \multicolumn{1}{l|}{4}          & \multicolumn{1}{l|}{0.20\%}                                                & \multicolumn{1}{l|}{7.16\%}                                                  & 12569          & 4.25                                                                         \\ \hline
\textbf{SL}                                                                          & \multicolumn{1}{l|}{2}          & 3708 & \multicolumn{1}{l|}{\textbf{1}}          & \textbf{4826} & \multicolumn{1}{l|}{\textbf{1}}          & \multicolumn{1}{l|}{4.44\%}                                                  & \textbf{9003} & \multicolumn{1}{l|}{\textbf{3}} & \multicolumn{1}{l|}{0\%}                                                  & 6401          & \multicolumn{1}{l|}{\textbf{1}} & \multicolumn{1}{l|}{4.73\%}                                                & \multicolumn{1}{l|}{8.90\%}                                                  & \textbf{17977} & 2.5                                                                          \\ \hline
\textbf{SS}                                                                          & \multicolumn{1}{l|}{4}          & 1782 & \multicolumn{1}{l|}{9}          & 1097          & \multicolumn{1}{l|}{8}          & \multicolumn{1}{l|}{4.41\%}                                                  & 4531          & \multicolumn{1}{l|}{10}         & \multicolumn{1}{l|}{0\%}                                                  & 2536          & \multicolumn{1}{l|}{10}         & \multicolumn{1}{l|}{2.73\%}                                                & \multicolumn{1}{l|}{3.64\%}                                                  & 8241           & 8.2                                                                          \\ \hline
\textbf{KG}                                                                          & \multicolumn{1}{l|}{6}          & 1551 & \multicolumn{1}{l|}{7}          & 1681          & \multicolumn{1}{l|}{10}         & \multicolumn{1}{l|}{0\%}                                                  & 3991          & \multicolumn{1}{l|}{8}          & \multicolumn{1}{l|}{0\%}                                                  & 3216          & \multicolumn{1}{l|}{8}          & \multicolumn{1}{l|}{1.50\%}                                                & \multicolumn{1}{l|}{4.99\%}                                                  & 10028          & 7.8                                                                          \\ \hline
\textbf{DD}                                                                          & \multicolumn{1}{l|}{\textbf{1}} & \textbf{3942} & \multicolumn{1}{l|}{5}          & 2635          & \multicolumn{1}{l|}{11}         & \multicolumn{1}{l|}{0\%}                                                  & 915           & \multicolumn{1}{l|}{11}         & \multicolumn{1}{l|}{0\%}                                                  & 2393          & \multicolumn{1}{l|}{11}         & \multicolumn{1}{l|}{1.33\%}                                                & \multicolumn{1}{l|}{1.33\%}                                                  & 7552           & 7.8                                                                          \\ \hline
\textbf{KP}                                                                          & \multicolumn{1}{l|}{9}          & 825  & \multicolumn{1}{l|}{6}          & 2597          & \multicolumn{1}{l|}{6}          & \multicolumn{1}{l|}{3.47\%}                                                  & 5760          & \multicolumn{1}{l|}{6}          & \multicolumn{1}{l|}{\textbf{10.24\%}}                                        & 3908          & \multicolumn{1}{l|}{\textbf{2}} & \multicolumn{1}{l|}{0.16\%}                                                & \multicolumn{1}{l|}{\textbf{12.25\%}}                                        & 15507          & 5.8                                                                          \\ \hline
\textbf{MH}                                                                          & \multicolumn{1}{l|}{8}          & 1051 & \multicolumn{1}{l|}{8}          & 1421          & \multicolumn{1}{l|}{5}          & \multicolumn{1}{l|}{0\%}                                                  & 5979          & \multicolumn{1}{l|}{5}          & \multicolumn{1}{l|}{1.83\%}                                                  & 5452          & \multicolumn{1}{l|}{\textbf{3}} & \multicolumn{1}{l|}{1.01\%}                                                & \multicolumn{1}{l|}{0\%}                                                  & 14795          & 5.8                                                                          \\ \hline
\textbf{TL}                                                                          & \multicolumn{1}{l|}{11}         & 366  & \multicolumn{1}{l|}{\textbf{3}} & 3412          & \multicolumn{1}{l|}{\textbf{3}} & \multicolumn{1}{l|}{0\%}                                                  & 8166          & \multicolumn{1}{l|}{4}          & \multicolumn{1}{l|}{0\%}                                                  & 5485          & \multicolumn{1}{l|}{9}          & \multicolumn{1}{l|}{1.87\%}                                                & \multicolumn{1}{l|}{2.14\%}                                                  & 9337           & 6                                                                            \\ \hline
\textbf{BF}                                                                          & \multicolumn{1}{l|}{10}         & 740  & \multicolumn{1}{l|}{\textbf{2}} & 4215          & \multicolumn{1}{l|}{7}          & \multicolumn{1}{l|}{9.14\%}                                                  & 5468          & \multicolumn{1}{l|}{9}          & \multicolumn{1}{l|}{0\%}                                                  & 2685          & \multicolumn{1}{l|}{7}          & \multicolumn{1}{l|}{\textbf{6.92\%}}                                       & \multicolumn{1}{l|}{2.87\%}                                                  & 10472          & 7                                                                            \\ \hline
\textbf{KF}                                                                          & \multicolumn{1}{l|}{3}          & 2889 & \multicolumn{1}{l|}{11}         & 282           & \multicolumn{1}{l|}{\textbf{2}} & \multicolumn{1}{l|}{\textbf{9.46\%}}                                         & 8445          & \multicolumn{1}{l|}{\textbf{2}} & \multicolumn{1}{l|}{0\%}                                                  & 8395          & \multicolumn{1}{l|}{5}          & \multicolumn{1}{l|}{\textbf{6.92\%}}                                       & \multicolumn{1}{l|}{1.62\%}                                                  & 12282          & 4.6                                                                          \\ \hline
\textbf{JE}                                                                          & \multicolumn{1}{l|}{7}          & 1492 & \multicolumn{1}{l|}{10}         & 470           & \multicolumn{1}{l|}{9}          & \multicolumn{1}{l|}{0\%}                                                  & 4250          & \multicolumn{1}{l|}{7}          & \multicolumn{1}{l|}{2.74\%}                                                  & 3651          & \multicolumn{1}{l|}{6}          & \multicolumn{1}{l|}{0.85\%}                                                & \multicolumn{1}{l|}{0\%}                                                  & 11753          & 7.8                                                                          \\ \hline
\end{tabular}
\Description[The rank and percent of points that each team scored for each of the five events.]{The rank and percent of points that each team scored for each of the five events. For most teams, collaboration increased from event three to five (2.81\% average to 4.08\%, with the highest-collaborating team going from 9.46\% to 12.25\%). Team SL consistently placed in the top three ranks, while Team KP improved their performance each event, going from rank 9 to 6 to 2.}
\caption{The rank and percent of points that each team scored for each of the five events. Percent collaboration
refers to the percent of total points that a team scored through collaboration, while percent tasks refer to the
percent of total points that a team scored by completing tasks.}
\label{tab:team_rank}
\end{table*}

Initially, DD-10 noted that ``no one does reporting flags either because that takes way more work to really put together a report.'' For Event 1, two reporting flags were submitted (0.88\% of all flags); and none were submitted for Event 2. After we increased the point values for submitting reporting flags for the third CTF, we found that students positively responded to this change. For Events 3, 4, and 5, the ratio of reporting flags to total flags increased to 5.45\%, 3.78\%, and 11\%, respectively. As a percentage of total points scored, this was 3.9\%, 2\%, and 5\% respectively.

Team KP and SL's strategies seemed to be prioritizing actions that would maximize the number of points scored. Along these lines, KP-15 said he largely focused on completing his own team’s flags. However, once we added the collaboration incentive, ``any flag that would pop up once I refreshed, I would just go for it, because we’re going to get those extra points by doing another team’s flags.'' In this way, Team KP consistently increased the number of points they obtained through collaboration, from 3.5\% and 10.3\% in Events 3 and 4, respectively, to 12.3\% in Event 5 (see Table~\ref{tab:team_rank}).

In Table~\ref{tab:team_rank}, for Event 5, we see a slight correlation with respect to how well teams ranked and whether they leveraged the collaboration and task features. For example, Team SL placed first, receiving 13.6\% of their points through collaboration and tasks, and Team KP placed second, receiving 12.4\% of points in a similar manner. However, Team MH did not obtain any points through collaboration and only 1\% of points through tasks, but still placed third.

\paragraph{Perceptions of fairness}
Multiple students, including OT-1, DD-10, DD-11, and KG-9, said that the balance of points improved over the course of the semester as we incorporated their feedback into the rubric. For instance, DD-10 said, ``Before, the more laborious tasks weren’t rewarded nearly as much as they should have been. I think now they are [rewarded] more.'' OT-1 added, ``I think originally archiving was 100 points \ldots To me that was way too much. And I think you guys lowered it to 50 or something now. So that makes sense, because archiving is easy, right?''

Still, we found opposing perspectives around how many points should be assigned to certain types of flags. KG-8 said that verification and reporting flags should be worth more points because they were more crucial to the investigation. %
In contrast, while OT-1 said that the balance of points for discovery, archival and reporting flags were fair, he believed that verification flags were worth too many points. This may be because OT-1, who has worked in the US Intelligence Community, said it’s ``really hard to say with 100\% probability [that something has been debunked].'' %

\subsubsection{Balancing Competition and Collaboration}
\label{sec:comp_collab}
As the semester progressed, we modified the point values in line with our goal to make CoSINT more collaborative, while also taking into account students’ feedback on the relative balance of points assigned to different flag categories. %
Still, we found a tension between competition and collaboration. Some students felt that collaboration was not incentivized enough for it to be worth the effort, or that it was unclear how they could collaborate with other teams. We conclude with ways to better incentivize collaboration.

\paragraph{Competitions promote efficiency and intra-team collaboration.}
We found that the competitive environment encouraged students to work more efficiently, both on their own and with others. While working individually, KP-14 said that CoSINT ``definitely enforced my thought process of, ‘How do I compete in a CTF? Where do I get my points from?’'' SS-4 and SS-5 felt similarly, saying, ``It helped us to understand the process overall, but it encouraged us to be more efficient in how we look at the process.''

\paragraph{Inter-team collaboration can be useful in competitions but is difficult to structure.}
Apart from encouraging competition and intra-team collaboration (within teams), we observed that CoSINT also promoted inter-team collaboration (between teams). Some students found the ability to collaborate with other teams useful, but others were unsure how to do so effectively.

We found that students appreciated the ability to gain points through collaboration for two reasons. First, this incentivized people to work together, and second, working on other teams’ flags gave them access to a wider variety of topics to investigate and methods to use. 

For example, when KG-8 and KG-9 learned that Team KF was performing well because they were using a Twitter scraping tool, they decided to look into it and have their team use it as well, because otherwise they thought that ``we’re totally going to get destroyed.'' KG-9 liked this balance between competition and collaboration in CoSINT where it is ``half collaboration and half competition.'' 

OT-1 also said that there were multiple instances where he was contributing to another team's evidence piece to gain points. By Event 5, Team OT obtained 7.36\% of their total points from collaboration and tasks. OT-1 also found leads from other teams that were beneficial for his own work. In one example, OT-1 said ``I was looking at another team’s discovery post, because I was going to archive it. And that account had over 100,000 followers, and I was like, ‘Well, I haven’t heard of this account before.’ \ldots So it was helpful to find other accounts through the CTF.''

Many students saw the ability to build on other teams' flags as a turning point in the semester. KP-14 recalled, ``As people got better at the CTF, they became more competitive and more collaborative. But as far as adding the feature of actually being able to verify other people’s stuff, that definitely had a significant boost on collaboration.'' KP-14’s team saw the collaboration feature as an ``opportunity to be a shark'' and earned many of their points in the final CTF this way. Specifically, KP-14’s team searched and filtered evidence tab for specific teams’ evidence pieces. Then, they inspected the status to determine if an archival, verification, or reporting flag was present. If there was a missing flag, someone from KP-14’s team would attempt to create it themselves. For Event 4, Team KP was the most collaborative team, obtaining 373\% more points through collaboration than the second-most collaborative team. For Event 5, Team KP placed second and was the second-most collaborative team at 12.41\% of total points, versus 13.63\% for Team SL who placed first (see Table~\ref{tab:team_rank}).

Some teams did not collaborate, and students described several reasons why not. First, SL-2 and DD-11 both said that collaborating with teams was not incentivized enough in the most recent version of CoSINT because it was not ``worth as much points as the time that went into doing them properly, [versus] making flags yourself'' (SL-2). Even though SL-2 said collaboration was not sufficiently incentivized, Team SL still obtained 13.63\% of their points through collaboration and tasks.

Second, some teams found collaboration was confusing, time-consuming, or required cognitively demanding tasks like context-switching and sensemaking of the other team's work products. SL-2 was ``not really sure how to collaborate through the CoSINT platform.'' DD-11 decided not to collaborate because ``there wasn’t really a lot of time to understand what the other teams were working on and what their objectives are, both from the flag and evidence perspective, and then from the tool perspective.'' SS-5 recalled, ``I haven’t helped another team’s flag yet. Even with the points I’m just not inclined to, because at the end of the day, I have to read through theirs, understand what it is. That’s almost like stopping in my tracks what I’m doing already, and then trying to understand what they’re doing.''

Third, some teams worried about how collaboration might negatively impact other teams. SS-4 and KP-15 voiced concerns that unexpected collaboration could be distracting:
\begin{quote} There’s a fine line between competition and collaboration with some things, because if I verify some other group’s piece of evidence, we're technically collaborating, but perhaps their strategy is to have them focus on their own pieces of evidence. So maybe I'm disrupting [their] strategy as well. (KP-15)\end{quote}

More broadly, some teams felt that the novelty of collaboration required changing norms or reframing expectations, especially for those with prior CTF experience. Although Teams OT, KG, and KP took advantage of collaboration features during the CTFs, even these teams recognized it as unusual. OT-1 said that collaboration was not common in typical OSINT investigations he had participated in. SS-4 thought the idea of collaborating across competing teams was a ``really cool idea and it definitely works [but] if all the teams are on board with doing that, it’ll go a lot smoother.''
OT-1 suggested a change in mindset for all participants at the CTF might help improve collaboration. Participants should view the CTF as teams collectively working towards a common goal ``instead of separate teams working on separate things, trying to win.''

\subsubsection{Balancing Breadth and Depth}
We described in Section~\ref{sec:design_goals} how one of the goals for CoSINT was to rapidly identify and debunk misinformation. This required casting a wide net, both in terms of covering a wide variety of topics but also collecting a large quantity of content. However, we found a tension between our design goals that emphasized breadth versus students’ desire to conduct in-depth investigations.

\paragraph{Defining a good investigation: breadth vs. depth.}
In Section~\ref{sec:comp_collab}, we described how the competitive environment promoted efficiency and breadth. However, OT-1, who had prior experience with OSINT investigations, believed that a good player needed to balance discovering a large quantity of content while making sure that it is also of high quality through careful research. He explained that ``a lot of times it’s easy to discover poor quality tweets made by bots, you know, it’s obvious, but then the real exploitable information is a little bit harder to find.''

Five other students also said that they preferred an environment that incentivized conducting in-depth investigations. For example, KG-8 said her team ended up finding ``a lot of small pieces of misinformation [because] a lot of the bigger fish had sort of been fried already.'' DD-10 also pointed out how his team had not spent much time looking into any single piece of evidence, but rather ``trying to just cast a huge net.'' He went on to describe what he perceived as the tension between breadth and depth within a competition: ``Stuff like that, that takes a lot of time, and quantitatively it’s not very much actual result at all, is actually the most [intrinsically] rewarding. You really have to be clever about this one image instead of finding all of them.''

\paragraph{Rewarding and assessing depth.}

Through our classroom deployment, we found that students were receptive to changes in the point structure. In turn, rewarding in-depth work with more points may satisfy some students’ desires to investigate in depth. For example, KP-15 said that they valued original verifications where they ``extrapolated on some knowledge from a few different sources \ldots as opposed to just using a fact check article,'' but also added that there was a big ``point boost'' for original verifications, which he described as a ``win-win'' scenario.

From Event 2, we began tracking if a verification flag identified an instance of \textit{misinformation} --- that is, whether it refuted the original claim made in the discovery flag. For example, one member of Team JE found a video posted online that claimed to show an instance of voter fraud. However, this team member was able to debunk the video by finding an alternative source that had investigated the same video. In this case, that verification flag identified an instance of misinformation. From Event 2 through Event 5, the percentage of approved verification flags with instances of misinformation rose from 31.8\% to 89.25\% (see Fig.~\ref{tab:evidence_flag}), indicating that teams increasingly submitted and investigated content fitting our definition of misinformation. (As a caveat, Event 5 focused on topics such as 9/11 and chemtrail conspiracy theories, which are more likely to contain misinformation.)

\subsection{Judging Improved Quality But May Be Difficult to Scale Up}

\subsubsection{Self-assessment rubric and judging improved quality of flags}
We found two aspects of the judging process that students said helped improve the quality of flags that they submitted. First, students perceived the self-assessment rubric as valuable. For example, KG-7 noted that the self-assessment rubric helped them better understand the requirements for a high-quality flag to ``make sure I can get the most points and I can justify the points.'' %

Second, students said that the judges’ feedback encouraged them to submit higher quality, more detailed flags that were not only more likely to be approved, but would be worth more points. Many of SS-5's discovery flags were rejected early on because he did not sufficiently describe why what he had found was potential misinformation. However after he resubmitted the same flags with more details, they were approved. %

\subsubsection{Judging misinformation may be subjective}
Despite our use of a rubric to make judging more fair and objective, three students --- DD-13, KP-14, and OT-1 --- pointed out that judging whether something is misinformation may be a subjective task. 
Further, KP-14 worried that the evidentiary standard required for verifying or refuting a claim differed between judges. He said that he would sometimes reject flags because he did not think that a student had submitted enough evidence, but was not sure if another judge would have rejected the flag for the same reason, perhaps because they were ``a little more timid to reject it.''

To overcome these challenges, KP-14 and KP-15 suggested rotating judges between teams. %
Alternatively, DD-13 suggested a tiered judging system where one judge would go over another judges’ evaluation, acknowledging the potential drawbacks of increased reviewing workload and confirmation bias among trusting judges. %

\subsubsection{Rubric enables judging to scale up}

The research team, including three to six research assistants, acted as judges for all of the CTFs. During the first two CTFs, we maintained our rate of evaluation in line with students’ rate of submission. Students soon became quicker and more adept at submitting flags, leading to two occasions where we could not evaluate all flags before class ended. For Events 3 and 4, to decrease judging bottlenecks and provide students with practice being a judge, we asked students from each team to sign up as a judge. With the students’ help, we found that judges were better able to keep up with the rate of submission. In Event 3, there were 25 judges who evaluated 292 flags, and in Event 4, there were 24 judges who evaluated 238 flags. In both events, judges took on average 10.6 minutes to judge a flag after it was submitted. For Event 5, we wanted students to fully participate in flag submission, so we recruited additional research assistants as judges. Here, 11 judges took 20.6 minutes on average to judge 597 flags. Thus, the judges were able to review four times as many flags, but the judging time per flag doubled. 

Two potential reason for this increased efficiency could be that students became more adept at submitting higher quality flags (that took less time to evaluate), but also that judges themselves became more adept at evaluating flags. We found that judges also perceived that the self-assessment rubric enabled them to evaluate flags more quickly. In KG-9's words, the rubric ``streamlines the process --- ‘Okay, well, I need to follow this link, I need to check all these things’ --- it makes it easier to approve it.'' SS-5 also felt that the rubric encouraged students to more accurately rate themselves, such that as a judge, he rarely adjusted the rating.

\section{Discussion}
In this work, we engaged in a four-month-long Research through Design (RtD) process to develop the CoSINT platform. We find that a RtD process helped us to improve and validate the design of the platform, moving closer towards our ideal preferred state described in our two design goals. 

First, we find that CoSINT enabled a rapid response to misinformation on social media by merging the complementary benefits of competition and collaboration. CoSINT's point system not only motivated teams to compete \textit{against} each other, but to also collaborate \textit{with} each other. Second, our findings show that CoSINT structured students’ work, allowing them to perform complex investigative tasks ranging from discovery and archiving to verification and reporting. CoSINT was flexible enough that students could investigate a range of topics, from COVID-19 and election misinformation to human rights violations and stock market rumors. 

Recall that we instantiated our two design goals using four of Dix’s heuristics for software appropriation \cite{dix_designing_2007}. However, Dix provides another important heuristic that we discuss next: \textit{learn from appropriation}. By observing how a system has been used and appropriated, we can redesign the system to better support users. Our mixed-methods evaluation allowed us to assess CoSINT against our two design goals, which we revisit below.

\subsection{Goal 1: Enable a Rapid Response to Misinformation on Social Media}
\subsubsection{CoSINT reduced inefficiencies compared to current CTF competitions.}
We showed that a crowd of 46 students could be motivated to quickly identify and debunk hundreds of pieces of potential misinformation in sessions as short as 60 minutes by creating a competitive environment with a points-based incentive structure. Building on prior work showing the benefits of competition in crowdsourcing \cite{belghith_compete_2022, tausczik_distributed_2018}, CoSINT demonstrated that traditionally theoretical CTFs can be adapted for real-world misinformation investigations. Second, we mitigated some key limitations of competitions, such as information silos, by allowing competing teams to view and build upon each others’ evidence and flags. In fact, some teams scored up to 12.25\% of their points by contributing to other teams' evidence pieces (average = 4.07\%), whereas such collaboration would not be possible in traditional CTFs.

Despite the noticeable increase in collaboration over the last three events for our CoCTF platform, our findings suggest that collaboration can be further encouraged. For example, some students indicated that they were hesitant to contribute to other teams' flags without explicit calls for help or social norms encouraging collaboration within the CTF. As Lessig posits in his New Chicago School theory \cite{lessig1999code}, there are four ways to regulate human behavior: laws, norms, markets, and architecture. While CoSINT leverages markets (extra points for collaboration) and architecture (information sharing, contributing flags to other teams' evidence, and tasks), future CoCTFs should explore how to frame policies and develop social norms to encourage collaboration. 
For example, sets of two competing teams could be required be physically or virtually co-located to minimize redundancy and maximize information sharing \cite{venkatagiri2021crowdsolve}. In terms of social norms, the expert could emphasize the shared goal that teams are working towards, and encourage members of different teams to build rapport with each other \cite{oren2011framework,liu1997norm}. The architecture of CoCTF systems could also facilitate social translucence \cite{erickson2000social} where teams can press a \textsf{help wanted} button to indicate that they are open to collaboration.

\subsubsection{CoSINT reduced inefficiencies compared to traditional crowdsourcing approaches in three ways.}
In many traditional crowdsourcing systems with monetary compensation, designers implement ``attention checks'' to make sure crowd workers are making an honest effort to complete the work. They also aggregate multiple, repetitive crowd inputs for the same microtasks to mitigate the effects of low-quality work or biases \cite{mason2012conducting}. Instead, CoSINT enabled high-quality work through a combination of a trusted and trained crowd \cite{mason2012conducting}, a point-based incentive system \cite{vonahn2008gwap}, self-assessment rubrics \cite{dow2012shepherding}, and real-time feedback from judges \cite{turkomatic}. Because we knew the students and developed a working relationship with them, we could trust them to submit higher-quality work compared to an anonymous crowd. We could also delineate and communicate low- and high-quality work to students through the point system and rubric. Finally, students felt that the self-assessment rubrics and judging mechanism improved their work in the short- and long-term.

Still, more structure within teams could lead to greater efficiency gains. For instance, some teams organically devised assembly line workflows and team leaders emerged over time; these teams frequently placed high on the leaderboard. In contrast, teams that employed free-for-all workflows with minimal collaboration among team members and no explicit team leader did not perform as well.

Our findings suggest that both types of teams may benefit from more explicit structure and roles, such as delineating the responsibilities for each team leader and assigning roles to each team member \cite{harris2019joining,venkatagiri2021crowdsolve,retelny2017noworkflow,luther2013leadership}. While Retelny et al.~\cite{retelny2017noworkflow} suggest that rigid workflows restrict adaptability, we find that too much freedom can hamper performance. Future work should explore providing flexible structures that teams can choose to use and modify based on their working styles. For example, the leader could mitigate unwanted redundancy by assigning team members to work on a specific topic or social media platform. To prevent judges from being overwhelmed by work, the leader could also conduct a preliminary evaluation of their flags before forwarding it to the judge. To further increase efficiency, individuals could be assigned or encouraged to focus on tasks that they preferred or excelled at, such as content discovery versus verification.

\bl{To increase the impact of CoCTFs, events can increase the overall number of participants and also involve a greater number of professional investigators to lead teams. However, as we learned in our four-month-long deployment, judges occasionally struggled to keep up with the rate of submissions as teams became more adept at creating flags. To address this bottleneck, designers could explore developing automated judging systems trained on past judges' evaluations \cite{kim2020toward} and provide participants with automated tailored feedback from large language models (LLMs) \cite{cao2023leveraging}.}

\subsection{Goal 2: Give the Crowd (More) Agency}
In traditional crowdsourcing systems, complex tasks are divided into microtasks that crowd workers complete independently, with little to no interaction with each other or agency in how to complete these tasks \cite{kittur2011crowdforge, bernstein2010soylent}. However, CoSINT builds on a growing body of literature that shows that crowds can perform more complex tasks, provided that they are sufficiently motivated and given adequate scaffolding, training, and agency \cite{retelny2017noworkflow, dow2012shepherding, harris2019joining}.

\subsubsection{Providing more agency can lead to a virtuous cycle.}
We also found that CoSINT helped students learn to more critically examine information online and develop a mental model for conducting investigations. This proved to be a virtuous cycle: between the first and fifth events, students submitted 65\% more evidence pieces and 163\% more flags, while maintaining flag approval ratings. In addition, students said that they enjoyed using CoSINT, possibly motivating them to continue participating in the events.

Students desired even greater agency to investigate topics in greater depth. While we designed CoSINT to provide a rapid response to misinformation (60--90 minutes), future work should study how to provide the crowd with greater agency and design longer-duration CoCTF events.

\bl{Participants in our study also noted that greater quantity did not always imply greater quality, and it may be beneficial for CoCTF organizers to empirically analyze the trade-offs between quantity and quality. Teams could be limited to a certain number of submissions per hour, or high-quality flags could be emphasized --- through point incentives and community norms \cite{lessig1998new} --- over low-quality flags.}

\subsubsection{Dynamically modify competition and collaboration affordances.} 
Though CoSINT motivated most students to participate, some were less motivated by competition, preferring collaboration instead. To better engage crowd workers, future CoCTF systems should consider alternative team structures and more flexible incentive and feedback mechanisms. 

Prior to starting the event, the system could survey the crowd to signal to the expert what motivates them \cite{pintrich2004conceptual}, and allow the expert to modify the system accordingly. One option would be to divide teams into two groups that work in the same environment but have different work arrangements. One group --- those motivated by a sense of urgency and competition --- could conduct rapid data collection and analysis with a group-wide competition or even self-competition \cite{michael_race_2020}. A second group --- motivated by conducting in-depth analyses and collaboration --- could collaboratively investigate the first groups' work in greater detail and over a longer period of time (days versus hours).

A second option would be to emphasize different types of feedback \cite{tekian2017qualitative}. For crowd workers who are more motivated by qualitative assessments of their work, the system could prompt judges to provide detailed written feedback, and emphasize this in the crowd worker's interface over the point values that the judge awarded them.

\subsection{\bl{Extending CoCTFs to Other Domains}}
\bl{Dix also recommends that designers \textit{allow for [re]interpretation} of the system. In other words, this intentional ``absence of meaning'' allows users to appropriate the system for other purposes \cite{gonzales2015appropriable}. As mentioned in Section~\ref{sec:related_work}, CTFs are best suited for settings focused on collecting new information or uncovering new problems \cite{karagiannis_analysis_2020}. Indeed, CoSINT could be easily adapted for domains other than online investigations of misinformation, such as to coordinate physical search-and-rescue efforts for missing persons or animals \cite{tracelabs_2022_twitter, white2014digital} or to assess damage after natural and man-made disasters \cite{bittner2016turning}. In these situations, flag types could remain the same (discovery, archival, verification, and reporting), but the evaluation criteria and point values for each would differ. For instance, the evaluation criteria for a discovery flag are currently originality, influence, and recency. These could be modified to focus on factors more relevant during crises, such as reliability and recency of information and level of danger posed. Given the potential consequences of errors when lives may be in immediate danger, organizers of such CoCTFs may also need to de-emphasize certain ``fun'' gamified and competitive elements \cite{susi2007serious}, and instead encourage greater communication and collaboration. 

To incorporate individuals with relevant skills and from diverse backgrounds, CoCTF organizers should consider working closely with existing communities of practice \cite{wenger1998cop} to understand their intrinsic and extrinsic motivations and enable novices to join these communities through legitimate peripheral participation \cite{lave1991lpp} For example, novices could be required to ``shadow'' judges or join an experienced team and make micro-contributions \cite{louie2021opportunistic}.}

\subsection{Limitations and Broader Impacts} Giving crowd workers more agency is often viewed in a positive light \cite{irani2013turkopticon}. In addition, CoSINT itself empowers regular citizens to work together to rapidly uncover misinformation, holding governments and corporations accountable for their words and actions. We must also grapple with the potential negative impacts that sociotechnical systems like CoSINT can have on individuals and societies \cite{hecht_its_2018}. For example, authoritarian governments could use CoSINT to crowdsource investigations into dissidents, or rogue crowds could use it to investigate members of marginalized communities. However, by incorporating expert supervision and evaluation, as well as training on professional and ethical investigative standards --- such as OSINT's ``no touch'' or passive reconnaissance ethos --- CoSINT reduces the likelihood of these potential harms.

\bl{In addition, addressing misinformation by delineating factual information from false and misleading information is not a panacea. Corrections may lead to a ``backfire'' effect and increase partisanship \cite{reinero2023factcheck}. Fact-checking may also not be relevant during crises and mass-convergence events --- such as natural disasters, protests, or political events --- where it may not be possible to immediately determine the veracity of information. In these situations, CoSINT can still be leveraged to improve investigators' contextual understanding of these events.} On the whole, we believe that \bl{by enabling democratic participation in understanding our online information ecosystems,} CoSINT can be used to do significantly more good than harm.

\section{Conclusion}

We engaged in a four month-long Research through Design process to develop and evaluate CoSINT, a platform for collaborative capture the flag competitions (CoCTFs) that enabled a trained crowd to investigate misinformation on social media. CoSINT showed that traditionally theoretical CTFs can be adapted for real-world misinformation investigations; and that novice crowd workers can be provided with greater agency in OSINT work when coupled with training, scaffolding, and expert guidance. Further, by incorporating beneficial elements of collaboration into a CTF, CoSINT ameliorates two limitations of purely competitive CTFs: information silos and duplication of effort. In turn, this novel CoCTF concept allowed a trained crowd of 46 students to identify and debunk hundreds of pieces of misinformation in less than ninety minutes, while collaborating up to 12\% of the time. By merging competition and collaboration, CoCTFs can be a powerful site of collective action that is both effective and enjoyable.

\begin{acks}
This paper about crowdsourcing was itself an effort in crowdsourcing. We would like to thank members of the Crowd Intelligence Lab for their contributions to this work --- especially Emily A., Ryan B., Charles C., Katie F., Alex H., Yunis H., Mariela J, Rissa M., Brandon N., Sophia P., and Raymar R. This work would also not have been possible without the involvement of students in the OSINT Lab course. We appreciate the insightful and detailed comments provided by Kate Starbird, Tanushree Mitra, Chris North, Eugenia Rho, and the anonymous reviewers. This work was supported by NSF award IIS-1651969 and the Virginia Commonwealth Cyber Initiative. Sukrit Venkatagiri was additionally supported by the University of Washington Center for an Informed Public, Craig Newmark Philanthropies, and the John S. and James L. Knight Foundation. Any opinions, findings, and conclusions or recommendations expressed in this material are those of the authors and do not necessarily reflect the views of the above supporting organizations or the National Science Foundation.
\end{acks}

\bibliographystyle{ACM-Reference-Format}
\bibliography{sample-base}

\appendix
\begin{figure*}[!htb]
\section{Appendix}
\label{appendix}

\hfill \break

\fbox{\includegraphics[width=7cm]{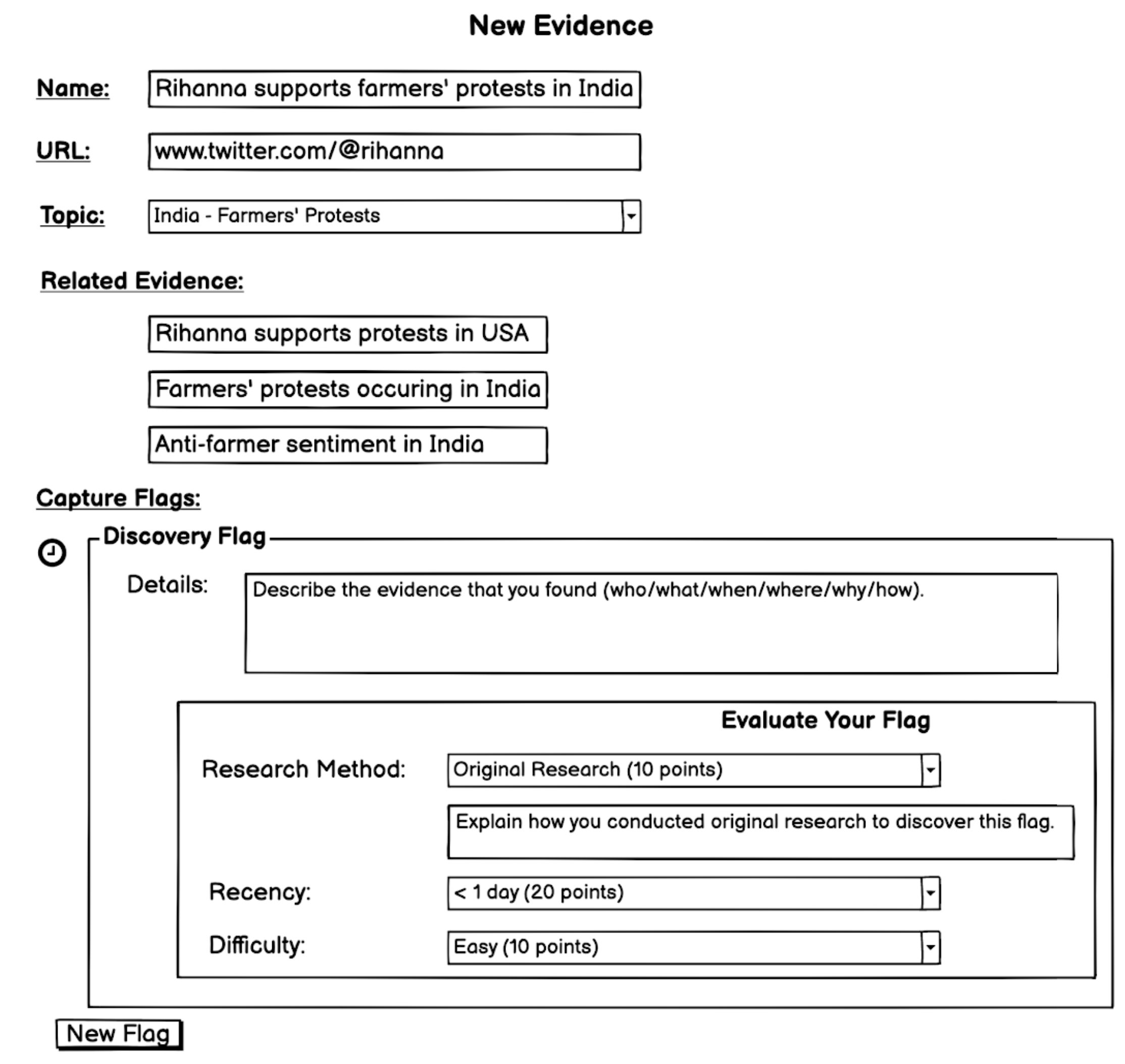}}
\Description[Low-fidelity wireframe example]{An example of an early low-fidelity wireframe that we created in Balsamiq. There are form options to enter the name, URL, and topic of the evidence piece. As well as three drop-down menus to choose related evidence pieces. Inset below, there is a form for users to enter details for their associated discovery flag, as well as a form to self-evaluate their flag (the method used, recency, and difficulty).}
\caption{An example of an early low-fidelity wireframe that we created in Balsamiq. This ``New Evidence'' page shows how a user would create a new evidence piece by specifying the name of the evidence, the URL for a social media post and what topic (now narrative thread) it is related to. This early version also allowed users to reference other evidence pieces to construct a broader narrative or story. At the bottom of the page, the user ``captures'' the discovery flag that awards them points for documenting this new evidence piece.}
\label{fig:low-fidelity-prototype}
\end{figure*}

\begin{figure*}[h]
\fbox{\includegraphics[width=10cm]{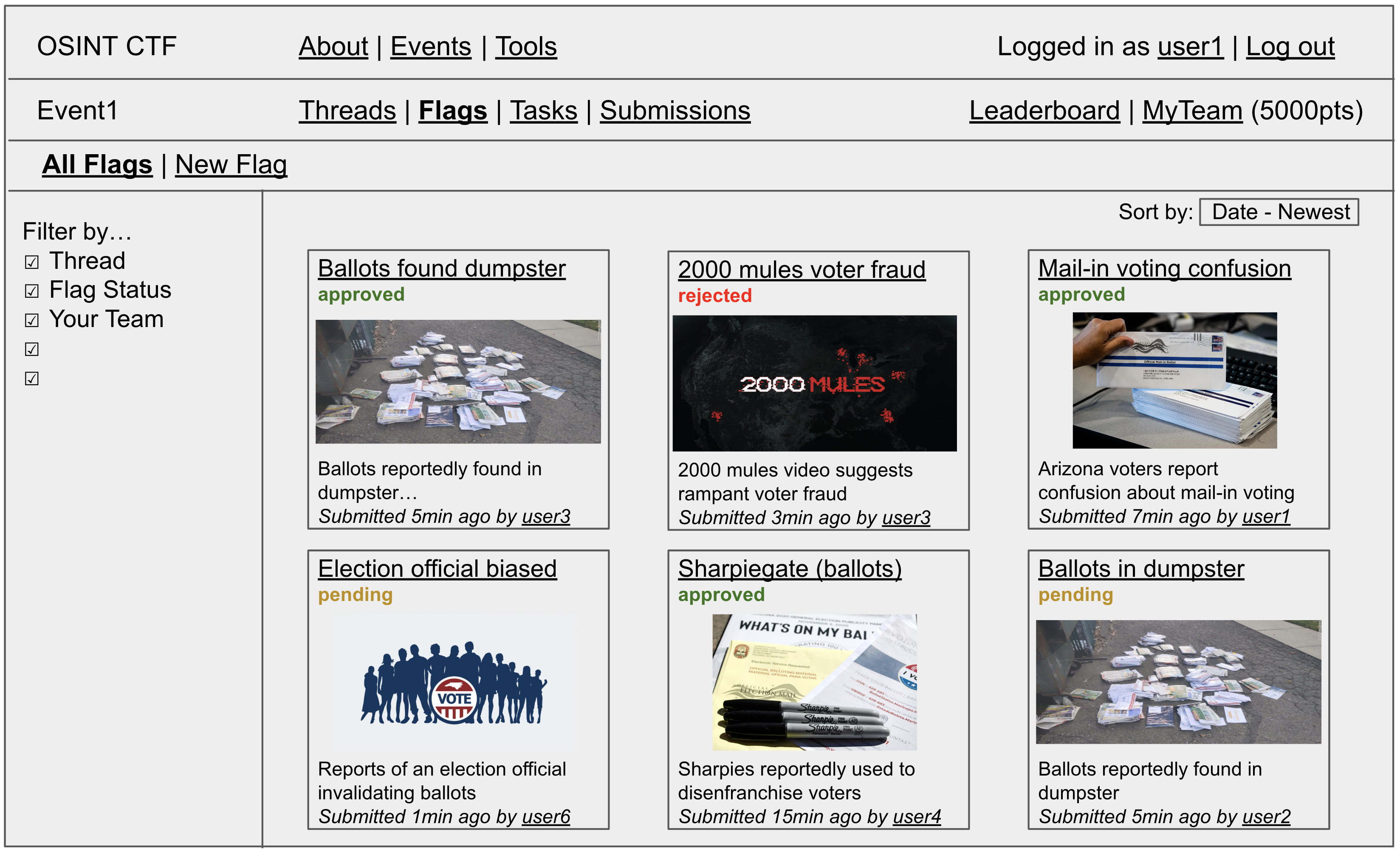}}
\Description[High-fidelity wireframe example]{An example of a high-fidelity wireframe that we created in Google Slides. This ``Flag List View'' page shows users a card-like interface of all flags submitted during an event (by all users). On the left are filters (for thread, its status, the team that created it, etc). On top are two menu bars that link to other pages (About, Events, Tools, Thread, Flags, Tasks, Submissions, Leaderboard). Inset is a 2x3 grid of cards, with each card representing a flag. The card contains the title of the flag, its approval status, a short description, when it was submitted, and which user submitted it.}
\caption{An example of a high-fidelity wireframe that we created in Google Slides. This ``Flag List View'' page shows users a card-like interface of all flags submitted during an event (by all users). On the left are filters allowing the user to narrow the set of flags based on the associated thread, its status, the team that created it, etc. On top are two menu bars that link to other pages. The topmost menu bar links to an about page, a page displaying all prior and future events, and all tools created by other users. The second menu bar links to pages that display narrative threads, flags, tasks, submissions (now renamed to evidence), the leaderboard, and the users' ``My Team'' page.}
\label{fig:high-fidelity-prototype}
\end{figure*}

\begin{figure*}[h]
\fbox{\includegraphics[width=12cm]{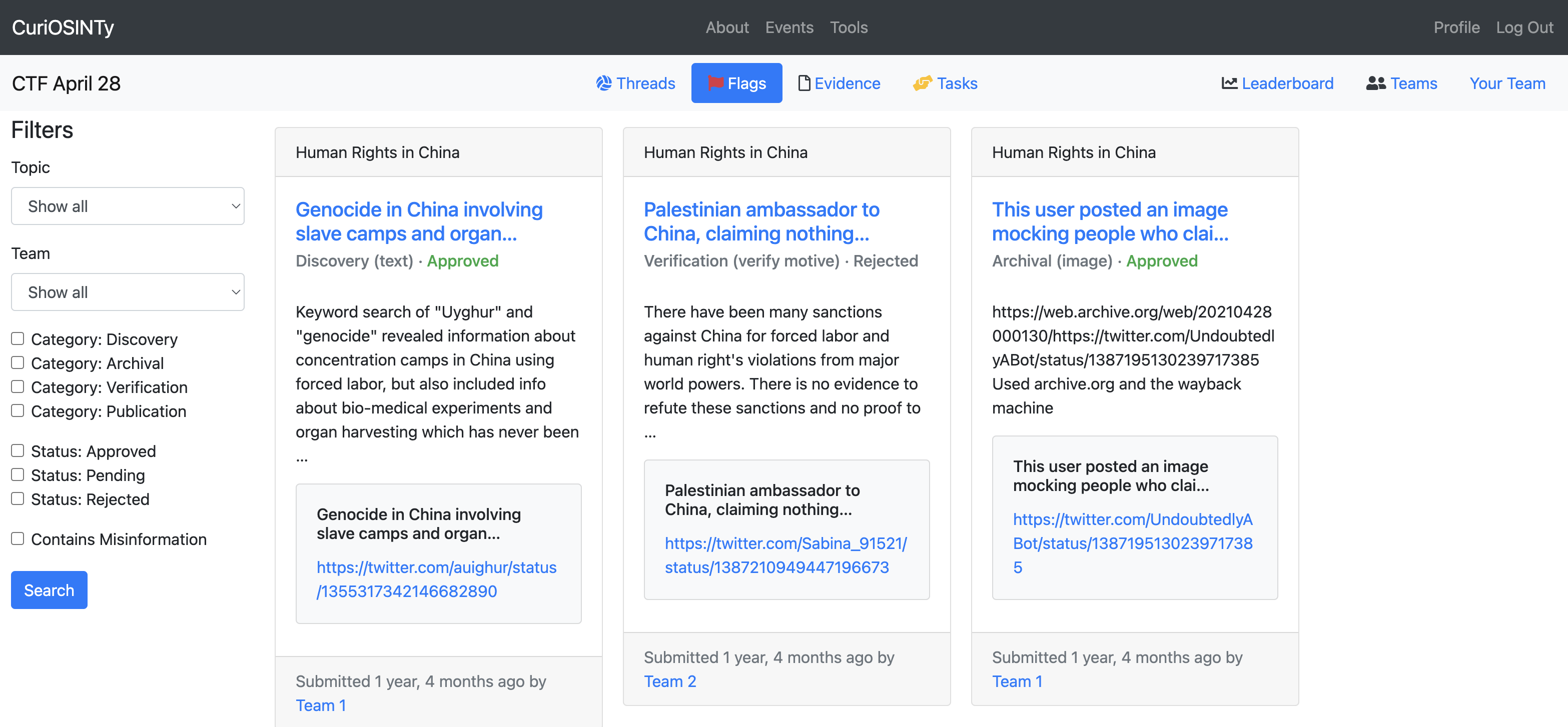}}
\Description[A screenshot of the actual CoSINT platform for the ``Flag List View.'']{A screenshot of the actual CoSINT platform for the ``Flag List View.'' This page shows users a card-like interface of all flags submitted during an event (by all users). Inset are the cards, each representing one flag. The card includes the name of the associated narrative, the title of the evidence it is part of, what type of flag it is, its status (approved, pending, or rejected), a description of the flag and the users' process of creating it, and the URL of the original discovery. On the left is a windowpane with filters that allow the user to narrow the set of flags based on the associated thread, its status, the team that created it, the type of flag, etc. On top are two menu bars that link to other pages. The topmost menu bar links to an about page, a page displaying all prior and future events, and a page listing all tools created by other users. The second menu bar links to pages that display narrative threads, flags, evidence, tasks, the leaderboard, a list of all teams, and the users' ``Your Team'' page.}
\caption{A screenshot of the actual CoSINT platform for the ``Flag List View.'' This page shows users a card-like interface of all flags submitted during an event (by all users).}
\label{fig:flag-view}
\end{figure*}

\begin{figure*}[h]
\fbox{\includegraphics[width=12cm]{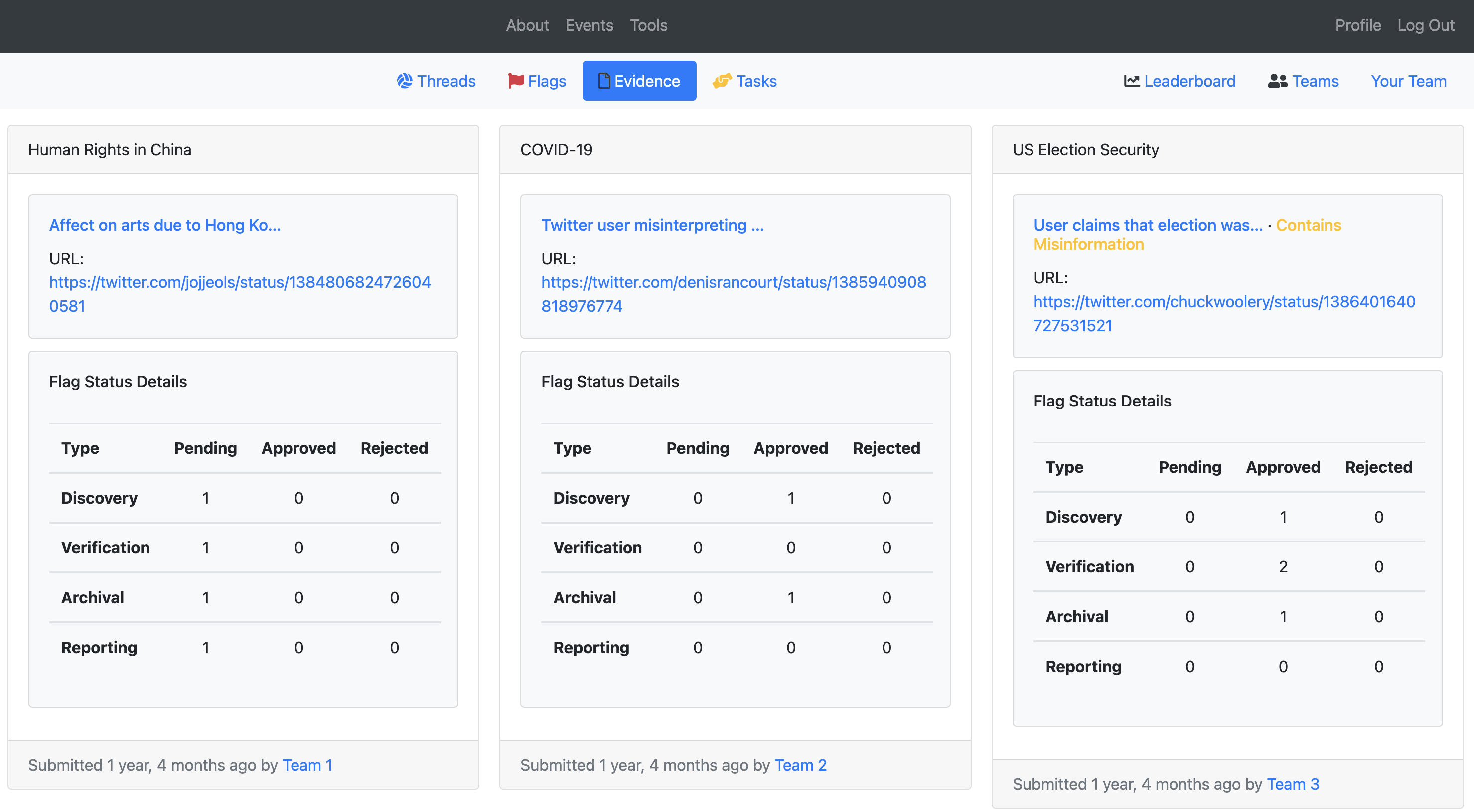}}
\Description[A screenshot of the actual CoSINT platform for the ``Evidence List View.'']{A screenshot of the actual CoSINT platform for the ``Evidence List View.'' This page shows users a card-like interface of all evidence pieces submitted during an event (by all users). Inset are the cards, each representing one evidence. The card includes a title, information about the original content found, and the status of all flag types that are part of that evidence --- whether and how many flags are pending, approved, or rejected. On the left (not pictured) are filters allowing the user to narrow the set of evidence pieces. On top are two menu bars that link to other pages. The topmost menu bar links to an about page, a page displaying all prior and future events, and a page listing all tools created by other users. The second menu bar links to pages that display narrative threads, flags, evidence, tasks, the leaderboard, a list of all teams, and the users' ``Your Team'' page.}
\caption{A screenshot of the actual CoSINT platform for the ``Evidence List View.''}
\label{fig:evidence-view}
\end{figure*}

\end{document}